\documentclass[journal]{IEEEtran}
\usepackage{amsmath,amssymb}
\usepackage{amsthm}
\usepackage{kotex}
\usepackage{amsfonts}

\usepackage{romannum}
\usepackage{graphicx}
\usepackage{caption}
\usepackage{subcaption}
\usepackage{breqn}
\usepackage{algorithm}
\usepackage{algpseudocode}
\usepackage{rotating}
\usepackage{pdflscape}
\usepackage{caption}
\usepackage{booktabs}
\usepackage{xcolor}
\usepackage{optidef}

\usepackage{tcolorbox}
\definecolor{RoyalPurple}{RGB}{120,81,169}

\definecolor{ForestGreen}{RGB}{34,139,34}

\definecolor{darkerlogocolor}{RGB}{20, 0, 145}  

\newtcolorbox{ttcolorbox}[1][]{colframe=darkerlogocolor, colback=darkerlogocolor!4!white, title=#1}

\algrenewcommand\algorithmicthen{}

\begin{document}
\title{\fontsize{22}{28}\selectfont Resilient LLM-Empowered Semantic MAC Protocols via Zero-Shot Adaptation and Knowledge Distillation}

\author{Yongjun~Kim,~\IEEEmembership{Graduate Student Member,~IEEE,}
            $^\dagger$Jihong~Park,~\IEEEmembership{Senior Member,~IEEE,}
        
$^\ddagger$Mehdi~Bennis,~\IEEEmembership{Fellow,~IEEE,}
        and~Junil~Choi,~\IEEEmembership{Senior Member,~IEEE}

\thanks{
Y. Kim and J. Choi are with the School of Electrical Engineering, KAIST,
Daejeon 34141, Korea (email: {yongjunkim, junil}@kaist.ac.kr).}
\thanks{$^\dagger$J. Park is with the Information Systems Technology and Design Pillar,
Singapore University of Technology and Design, Singapore 487372 (email:
jihong\_park@sutd.edu.sg).}
\thanks{$^\ddagger$M. Bennis is with the Centre for Wireless Communications, University of Oulu, Oulu 90570, Finland (email: mehdi.bennis@oulu.fi).
}
\thanks{J. Choi and J. Park are corresponding authors.}
}


\maketitle

\begin{abstract}
Neural network-based medium access control (MAC) protocol models (NPMs) improve goodput through site-specific operations but are vulnerable to shifts from their training network environments, such as changes in the number of user equipments (UEs) severely degrade goodput. To enhance resilience against such environmental shifts, we propose three novel semantic MAC protocol frameworks empowered by large language models (LLMs). First, we introduce a token-based protocol model (TPM), where an LLM generates MAC signaling messages. By editing LLM instruction prompts, TPM enables instant adaptation, which can be further enhanced by TextGrad, an LLM-based automated prompt optimizer. TPM inference is fast but coarse due to the lack of real interactions with the changed environment, and computationally intensive due to the large size of the LLM. To improve goodput and computation efficiency, we develop T2NPM, which transfers and augments TPM knowledge into an NPM via knowledge distillation (KD). Integrating TPM and T2NPM, we propose T3NPM, which employs TPM in the early phase and switches to T2NPM later. To optimize this phase switching, we design a novel metric of meta-resilience, which quantifies resilience to unknown target goodput after environmental shifts. Simulations corroborate that T3NPM achieves 20.56\% higher meta-resilience than NPM with 19.8$\times$ lower computation cost than TPM in FLOPS.
\end{abstract}


%
\IEEEpeerreviewmaketitle

\section{Introduction}
In traditional communication systems, control messages for medium access control (MAC) protocols have been standardized and remain unchanged across diverse environments. This approach ensures compatibility but limits adaptability to varying traffic patterns and application requirements. As 6G takes shape with unforeseen and complex use cases, there is growing interest in tailoring MAC control messages to accommodate site-specific traffic patterns and applications. For instance, in low-traffic sites, uplink latency can be significantly reduced by eliminating unnecessary information in control messages, such as buffer status reports (BSRs) within protocol data units (PDUs) \cite{Valcarce:2024}.

To enable site-specific MAC operations, data-driven MAC protocols have emerged as a promising solution. One notable framework is the protocol learning framework, also known as the neural protocol model (NPM) \cite{Mota:2021}. NPM is built on multi-agent deep reinforcement learning (MADRL), wherein user equipments (UEs) are mapped into multiple agents that communicate control messages through a set of shared neural network (NN) layers associated with their base station (BS). By training the MADRL in specific environments and applications, NPM dynamically learns efficient control messages, leading to significant improvements in communication efficiency, such as higher goodputs and lower collision rates compared to traditional MAC protocols \cite{Seo:2023}.

\begin{figure}[t]
\centering
\includegraphics[width=0.86\columnwidth]{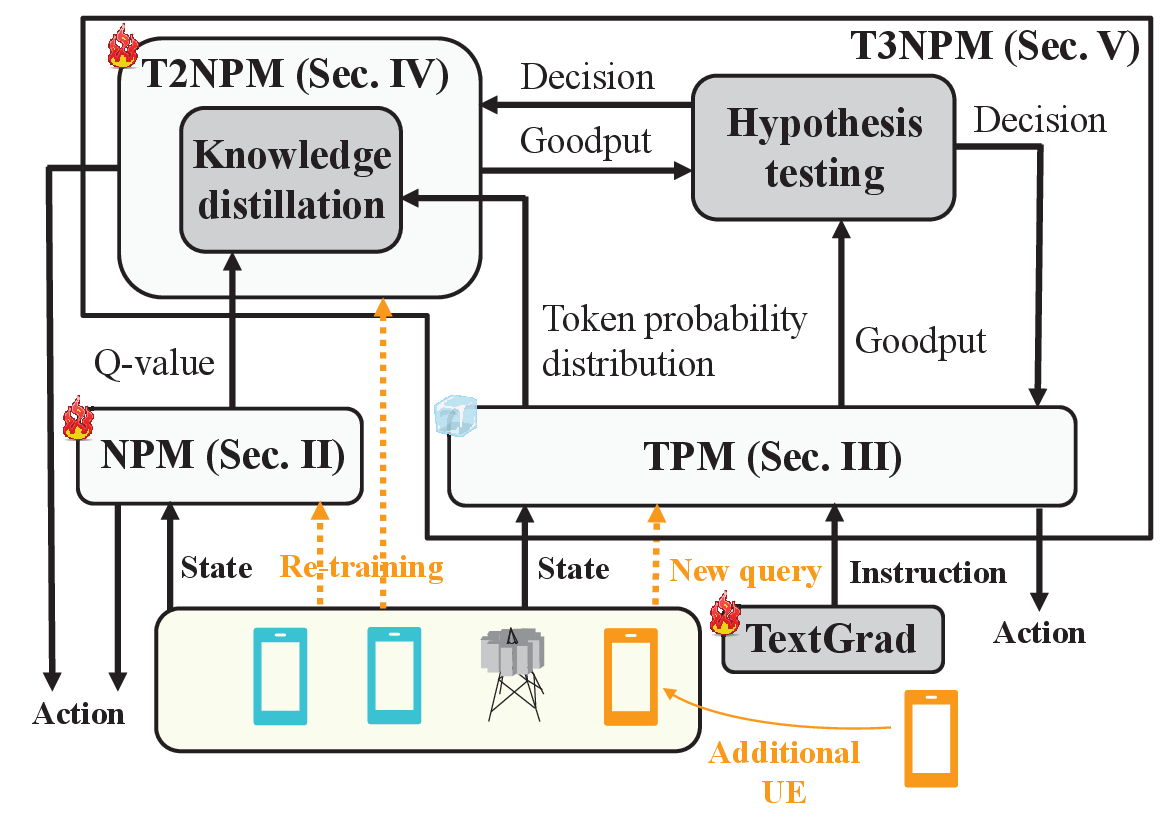}
\caption{LLM-empowered semantic MAC protocols.}
\vspace{-1em} 
\label{fig:comparing methods}
\end{figure}

Despite its promise, NPM faces significant challenges, especially from network environmental shifts between training and actual testing due to volatile factors like node mobility, node density, and more \cite{Hussien:2023}. 
For example, a critical issue arises when the number of UEs exceeds the original training scenario. Since UEs share BS NN layers in NPM, increase in UE numbers requires an expansion of BS NN dimensions, introducing untrained parameters. This necessitates frequent re-training using MADRL, thereby reducing NPM resilience.

To address the compromised resilience of NPM, as illustrated in Fig.~\ref{fig:comparing methods}, we first introduce a semantic MAC protocol framework leveraging large language models (LLMs), referred to as the token-based protocol model (TPM). In TPM, UE exchanges control messages using natural language tokens, with the LLM serving as the BS. Unlike NPM, which depends on an NN architecture trained via MADRL, TPM generates decisions through LLM inference based on its general knowledge, allowing for immediate responses without the need for re-training. However, since this general knowledge is not tailored to actual environments, TPM’s responses are typically less accurate than those of a re-trained NPM. Additionally, the substantial size of the LLM results in significantly higher computational costs for TPM compared to a re-trained NPM.

Considering these trade-offs between TPM and NPM, we highlight a bifold role for LLMs in data-driven MAC: 1) a first responder, providing immediate responses via TPM while awaiting NPM re-training; and 2) a teacher, guiding the re-training process through knowledge distillation (KD) from TPM-to-NPM (T2NPM). Accordingly, we integrate these two roles in the TPM-after-T2NPM (T3NPM) framework, wherein TPM is used initially and switches to T2NPM for enhanced performance. This new paradigm underscores the transformative potential of combining neural and semantic approaches to MAC protocols, paving the way for more adaptive and efficient communication systems in the 6G era. 


\subsection{Related Works: Protocol Learning and LLM for Commun.}

Data-driven MAC protocol design, particularly through reinforcement learning (RL), has enabled site-specific protocol optimization in complex network environments, which is difficult to achieve with traditional hand-crafted and standardized MAC protocols. For instance, prior studies have demonstrated the joint optimization of channel access, rate adaptation, and channel switching \cite{Xin:2024}, as well as the optimization of backoff types and modulation and coding schemes (MCS) \cite{Keshtiarast:2024}. 
Beyond these efforts focused on RL-based MAC configuration and policy optimization with fixed control signaling messages, recent advances on data-driven MAC protocols have further enabled site-specific and goal-oriented control signaling messages by learning them through MADRL \cite{Mota:2021, Naeem:2024}, often referred to as MAC protocol learning or NPM. In NPM, control messages are initialized as arbitrary representations within the NN layers of MADRL architecture, and gradually evolve into meaningful forms through the MADRL training process. These learned control messages can be extracted and structured using probabilistic logic in symbolic MAC, enhancing interpretability and computational efficiency \cite{Seo:2023}. However, existing MADRL-based MAC protocols typically assume stationary network environments so are vulnerable to environmental shifts such as changes in the number of UEs. These shifts necessitate re-training, revealing a critical resilience issue that, to our knowledge, has not yet been addressed.

On the other hand, recent advances in LLMs have led to their adoption to various layers in wireless communication systems. In the physical (PHY) layer, the frameworks proposed in \cite{Jiang:2024, Bariah:2023} transform multimodal data into textual forms to enhance personalization and beamforming performance. In addition, LLM's prediction capability has been utilized to predict future channels and mobility to proactively optimize transmit beams \cite{Sheng:2025} and movable antenna configurations \cite{Zhang:2025}. In the network and application layers, LLM's rich knowledge and general inference capability have been leveraged for flexible optimization of network slicing \cite{Lotfi:,Wu:2025} and service management and orchestration (SMO) with broadest network control \cite{Motalleb:2024}. However, LLM applications to the MAC layer remain limited. One notable study \cite{Tan:2025} fine-tunes an LLM-based MAC protocol via RL to address environmental shifts, but at the cost of thousands of training steps due to the large LLM model size. By contrast, in this paper, we explore the role of LLMs to accelerate the fine-tuning of MADRL-based MAC protocols and to enable switching from LLM-based to MADRL-based MAC protocols, thereby enhancing resilience in real-time MAC protocol operations.

\subsection{Contributions and Paper organization}
\label{subsection:contribution}
The main contributions of this paper are summarized below:
\begin{enumerate}
    \item 
    We propose TPM, which leverages an LLM to immediately adapt to environmental shifts, unlike NPM that requires re-training. To enhance TPM's understanding of shifted environments, we employ TextGrad, which automatically fine-tunes instruction prompts through iterative LLM self-reflection.

    \item 
    To provide more precise adaptation with reduced re-training and computational costs, we introduce T2NPM, which utilizes TPM to accelerate NPM re-training via KD. As the re-trained NPM is much smaller than the LLM-based TPM, T2NPM improves memory and computation efficiency.

    \item 
    To combine the respective benefits of TPM and T2NPM, we propose T3NPM, which initially applies TPM, and then switches to T2NPM once it surpasses TPM. Accurate switching requires multiple performance measurements. To reduce additional measurement costs, we develop MixSwitch, which determines switching based on a mixture of new and previous measurements.
    
    \item 
    Finally, we design a new metric, meta-resilience, which measures average resilience across all target performance levels, unlike existing resilience metrics tied to a fixed target level. This allows resilience to be quantified under environmental shifts where target levels are unknown, enabling optimization of T3NPM and comparison across our proposed semantic MAC protocols as well as with other baselines. Simulations show that T3NPM achieves highest meta-resilience, outperforming slotted ALOHA (S-ALOHA) by $23.53\%$ and NPM by $20.56\%$, while reducing computation cost by 19.8$\times$ compared to TPM in floating-point operations per second (FLOPS).
\end{enumerate}

The rest of the paper is organized as follows. Sec.~\ref{section:system model} introduces the system model, including NPM, which serves as a baseline. Sec.~\ref{section:TPM} presents TPM, especially using TextGrad for automatic instruction design. Sec.~\ref{section:T2NPM} describes T2NPM, which employs TPM as a teacher model and NPM as a student model during the training phase. Sec.~\ref{section:T3NPM} introduces T3NPM, which integrates TPM and T2NPM, and presents the proposed meta-resilience metric. Sec.~\ref{section:simulation} provides simulation results for the proposed methods using the meta-resilience metric, and Sec.~\ref{section:conclusion} concludes the paper.

\begin{figure}[t]
\centering
\includegraphics[width=0.9\columnwidth]{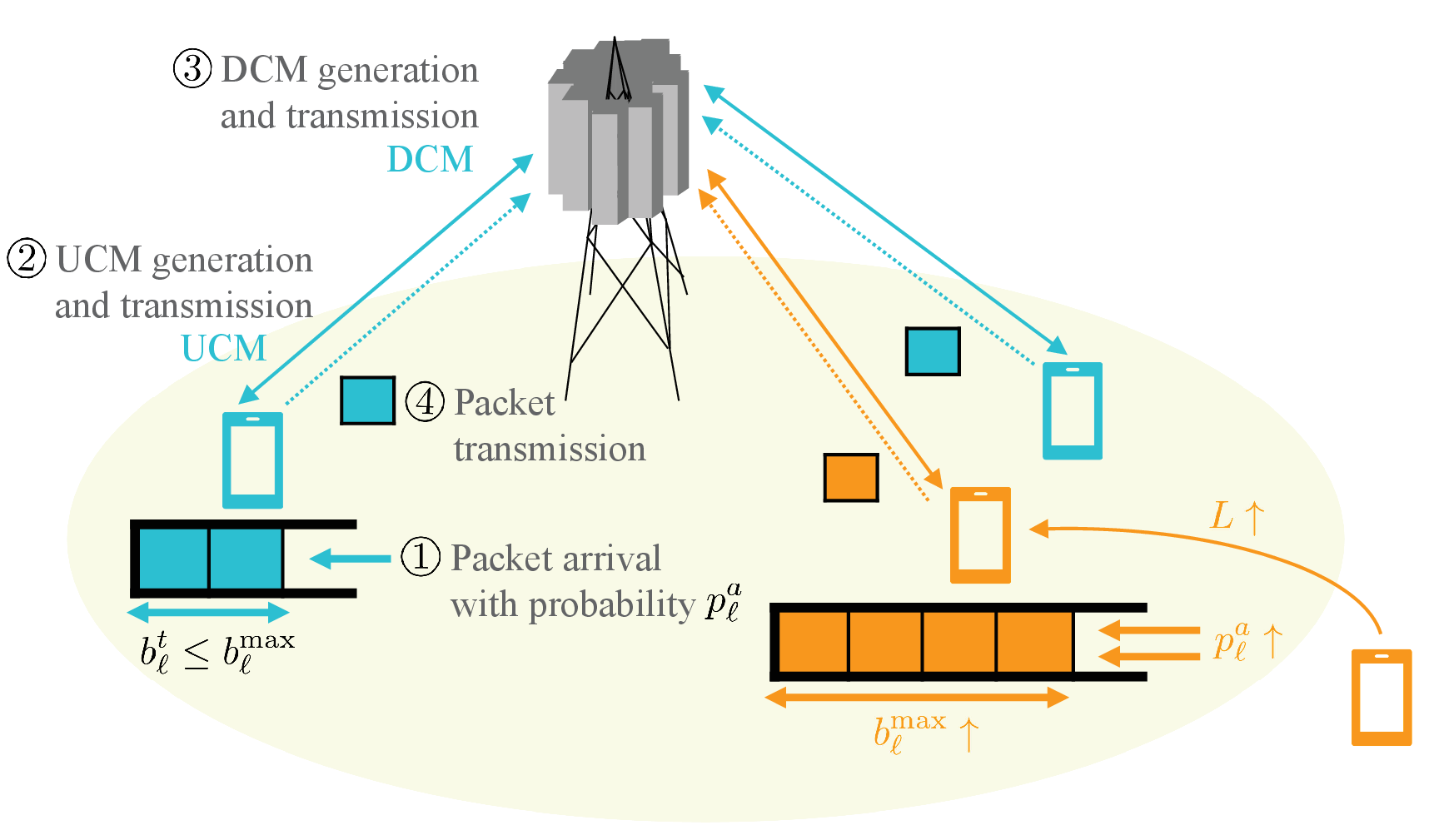}
\caption{Network model under environmental shifts.}
\vspace{-1.5em} 
\label{fig:system model}
\end{figure}

\section{System Model}
\label{section:system model}
\subsection{Multiple Access Network Channel}
\label{subsection:multiple access network channel}
Consider an uplink multiple access network consisting of a single BS and $L$ UEs. The network operates in discrete time slots $t = 0, 1, \cdots$ with a constant interval defined by a transmission time interval (TTI). At the beginning of time slot $t$, each UE $\ell$, where $\ell \in [1, L]$, has a buffer $\mathcal{B}_\ell^{t}$ that stores incoming packets. Following the first-in-first-out (FIFO) rule, each UE independently decides whether to transmit a packet from $\mathcal{B}_\ell^{t}$ over the uplink shared channel (UL-SCH), discard the packet, or remain silent. The BS successfully receives a transmitted packet if 1) there is no collision among UEs and 2) the packet is successfully decoded, where the UL-SCH is modeled as a packet erasure channel with a block error rate (BLER) $p_{\ell}^e$. To avoid collisions, UEs and the BS exchange control messages over a separate control channel prior to packet transmissions, as illustrated in Fig.~\ref{fig:system model}.

Specifically, at the beginning of time slot $t$, a packet arrives at $\mathcal{B}_\ell^{t}$ according to a Bernoulli process with arrival probability $p_{\ell}^a$. The buffer of UE $\ell$ can hold up to $b_{\ell}^{\max}$ packets. Defining the current number of packets in the UE $\ell$'s buffer as $b_{\ell}^t$, when the buffer overflows, i.e., $b_\ell^{t} > b_{\ell}^{\max}$, the oldest packet is discarded based on the FIFO rule. Each UE observes its $b_{\ell}^{t}$ and constructs an uplink control message (UCM). The BS receives the UCMs from all UEs and additionally observes the UL-SCH state $b_{0}^{t}$, which is defined as follows: the BS successfully receives a packet from UE $\ell$ ($b_{0}^{t} = \ell$); the BS receives no packets because all UEs are silent (i.e., idle UL-SCH, $b_{0}^{t} = 0$); or the BS fails to receive any packets due to packet erasure and/or collision events ($b_{0}^{t} = L+1$). Based on the received UCMs and the UL-SCH observation, the BS constructs and transmits downlink control messages (DCMs) to individual UEs. Consequently, each UE $\ell$ receives its DCM and takes an action $a_\ell^{t} \in \mathcal{A}$, where $\mathcal{A} = \{0, 1, 2\}$ denotes the three possible actions: being silent ($a_\ell^{t} = 0$), transmitting a packet ($a_\ell^{t} = 1$), or discarding a packet ($a_\ell^{t} = 2$).

To model network environmental shifts, we consider sudden changes in the environment parameters $p_{\ell}^a$, $b_{\ell}^{\max}$, and/or $L$, as illustrated in Fig.~\ref{fig:system model}. Unless otherwise stated, we hereafter focus solely on the shift caused by an increase in $L$. In NN-based protocol learning, such a shift results in an increased NN model size, thereby requiring re-training and significantly reducing system resiliency. To isolate the impact of this resiliency issue on UCMs and DCMs, we assume, for simplicity and without loss of generality, that the control channels for UCMs and DCMs are error-free.


\begin{figure}[t]
\centering
\includegraphics[width=0.85\columnwidth]{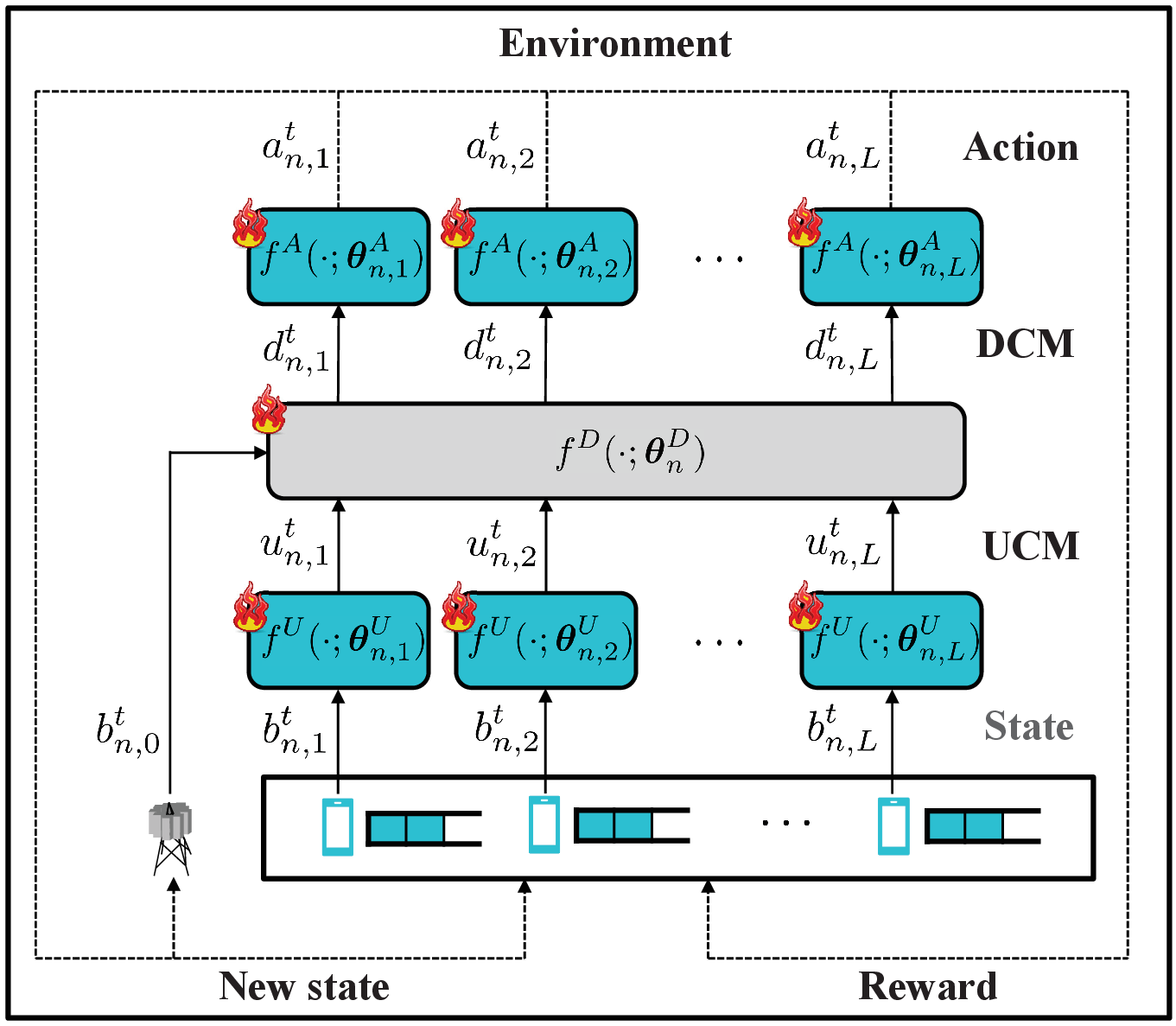}
\caption{A schematic illustration of NPM.}
\vspace{-1em} 
\label{fig:NPM}
\end{figure}

\subsection{NPM Operation}
\label{subsection:NPM operation}
We first introduce NPM as a baseline protocol learning framework. Following \cite{Mota:2021}, NPM jointly trains UCMs, DCMs, and UE actions using MADRL. For simplicity, instead of employing multi-agent deep deterministic policy gradient (MADDPG) in \cite{Mota:2021}, we follow \cite{Seo:2023} and adopt independent deep Q-learning (IDQN) \cite{Tampuu:2017} as the MADRL method. After training the model for $T$ TTIs, the trained model is deployed during the testing phase. We denote this interval as a single episode $n \in \mathcal{N}$ consisting of $T$ TTIs, where $\mathcal{N} = \{n_0, n_0 + 1, \ldots, \hat{n}\}$ represents the set of episode indices. Each episode $n$ is associated with its initial and terminal states, as defined in \cite{Sutton:2018}. At the initial state, the UL-SCH is idle, and both BS and UE buffers are empty. The terminal state corresponds to the buffer and channel observations at time $T$. Hereafter, the subscript $n$ denotes values corresponding to episode $n$.

The architecture of NPM is visually summarized in Fig.~\ref{fig:NPM}. 
The functions $f^U$, $f^D$, and $f^A$ in NPM represent NNs used to construct UCMs, DCMs, and Q-values of actions, respectively. For UE $\ell$, the UCM and Q-values are generated using $f^U$ and $f^A$, parameterized by $\boldsymbol{\theta}_{n,\ell}^U$ and $\boldsymbol{\theta}_{n,\ell}^A$, respectively. For constructing DCMs at the BS, $f^D$ is parameterized by $\boldsymbol{\theta}_n^D$. Aggregating these parameters across all UEs and the BS, we define the NPM parameter set at episode $n$ as $\boldsymbol{\Theta}_n = \left\{ \bigcup_{\ell=1}^L \boldsymbol{\theta}_{n,\ell}^{U},\ \boldsymbol{\theta}_{n}^{D},\ \bigcup_{\ell=1}^L \boldsymbol{\theta}_{n,\ell}^{A} \right\}.$ At episode $n$ and time slot $t$, each UE $\ell$ constructs its UCM as $u_{n,\ell}^{t} = f^U\left(b_{n,\ell}^{t}; \boldsymbol{\theta}_{n,\ell}^{U}\right),$ based on its buffer observation $b_{n,\ell}^{t}$. The BS then constructs DCMs for all UEs using its UL-SCH state $b_{n,0}^{t}$ and the received UCMs as $\bigcup_{\ell=1}^{L} d_{n,\ell}^{t} = f^D\left(\bigcup_{\ell=1}^{L} u_{n,\ell}^{t}, b_{n,0}^{t}; \boldsymbol{\theta}_{n}^{D}\right)$, where $d_{n,\ell}^{t}$ is the DCM intended for UE $\ell$. After receiving its DCM, each UE $\ell$ selects an action $a_{n,\ell}^{t}$ by choosing the one with the highest Q-value: $a_{n,\ell}^{t} = \arg\max_{a \in \mathcal{A}} f^A\left(d_{n,\ell}^{t}; \boldsymbol{\theta}_{n,\ell}^{A}\right).$ The pre-trained parameter set before the environmental shift is denoted by $\boldsymbol{\Theta}_0$, which is subsequently re-trained over episodes until $\hat{n}$.

Given $\boldsymbol{\Theta}_n$, the goodput is evaluated as the performance during the testing phase. Goodput is defined as the average rate of successful packet reception over $T$, i.e.,
\begin{equation}
G_n = \frac{\sum_{t=0}^{T-1} c_{n}^{t}}{T} \quad \text{(packets/TTIs)},
\label{eq:goodput}
\end{equation}
where $c_{n}^{t} = 1$ if the BS successfully decodes the arrived packet at time $t$, and $c_{n}^{t} = 0$ otherwise. The latter includes cases of collision, packet erasure, idle UL-SCH, or duplicated packet decoding (i.e., when the BS decodes the packet that has already been successfully decoded in previous time slots).

\subsection{NPM Training}
\label{subsection: NPM training}
While a trained NPM operates at the single TTI during the testing phase, the unit time interval during training is defined as $B \gg 1$ TTIs, since training computations typically incur significantly larger delays than communication operations. Let the discrete training time slot be denoted by $t' = 0, 1, \cdots$, with each step corresponding to an interval of $B$ TTIs. During training, the NPM parameter set is denoted by $\boldsymbol{\Theta}_n^{t'} = \left\{ \bigcup_{\ell=1}^L \boldsymbol{\theta}_{n,\ell}^{U,t'},\ \boldsymbol{\theta}_{n}^{D,t'},\ \bigcup_{\ell=1}^L \boldsymbol{\theta}_{n,\ell}^{A,t'} \right\},$ which is optimized over $t'$ and $n$ to maximize the goodput while avoiding collisions and unintended packet discards.

At UE $\ell$, the action $a_{n,\ell}^{t'}$ is selected using an $\epsilon$-greedy policy to balance exploration and exploitation \cite{Sutton:2018} as
\begin{align}
    a_{n,\ell}^{t'} = \arg\max_{a \in \mathcal{A}} f^A\left(d_{n,\ell}^{t'}; \boldsymbol{\theta}_{n,\ell}^{A,t'}\right) \cdot \mathbf{1}_{\eta_{\ell}^A = 0} 
    + \tilde{a} \cdot \mathbf{1}_{\eta_{\ell}^A = 1},
    \label{eq:epsilon-greedy}
\end{align}
where $\tilde{a} \sim \text{Unif}\left(\mathcal{A}\right)$ is an action sampled uniformly at random from the action space $\mathcal{A}$, and $\mathbf{1}_{\delta}$ is an indicator function that returns $1$ if the condition $\delta$ is true, and $0$ otherwise. According to the $\epsilon$-greedy policy, $\eta_{\ell}^A = 0$ with probability $1 - \epsilon$, in which case the greedy action is selected. Conversely, with probability $\epsilon$, we have $\eta_{\ell}^A = 1$, and a random action $\tilde{a}$ is selected to encourage exploration.

After executing all UEs' actions, the reward that UE $\ell$ receives is defined as
\begin{align}
    r_{n,\ell}^{t'} = \begin{cases}
    +\rho_1 , & \text{if } c_{n}^{t'} = 1, \\
    +\rho_2 , & \text{if } a_{n,\ell}^{t'} = 2 \text{ and } \mathcal{B}_{n,\ell}^{t'} \backslash \mathcal{B}_{n,\ell}^{t'+1} \in \mathcal{B}_{n,0}^{t'}, \\
    -\rho_3 , & \text{if } a_{n,\ell}^{t'} = 2 \text{ and } \mathcal{B}_{n,\ell}^{t'} \backslash \mathcal{B}_{n,\ell}^{t'+1} \notin \mathcal{B}_{n,0}^{t'}, \\
    -\rho_4 , & \text{if } \sum_{\ell'=1}^L \mathbf{1}_{\eta_{\ell'}^e = 0} \cdot \mathbf{1}_{a_{n,\ell'}^{t'} = 1} > 1, \\
    -\rho_5 , & \text{otherwise}, \\
    \end{cases}
    \label{eq:reward}
\end{align}
where $\mathcal{B}_{n,0}^{t'}$ denotes the set of packets decoded at the BS, and $\eta_{\ell}^e$ expresses whether the packet that UE $\ell$ transmitted is erased ($\eta_{\ell}^e = 1$), or not ($\eta_{\ell}^e = 0$) in UL-SCH modeled as packet erasure channel. When the BS successfully decodes a new packet, directly contributing to goodput, a reward of $+\rho_1$ is given to the UE that transmitted the packet. Other reward terms influence goodput indirectly. If UE $\ell$ discards a packet that has already been decoded at the BS, the UE receives a reward of $+\rho_2$, as this action helps the UE free buffer and enables reception of new packets in subsequent time slots. Conversely, if UE $\ell$ discards a packet that has not yet been decoded at the BS, a penalty of $-\rho_3$ is given due to the resulting packet loss and the missed transmission opportunity. In the event of a collision, i.e., when more than one UE transmits simultaneously, UE $\ell$ receives a penalty of $-\rho_4$ if it participates in such a collision. Finally, a penalty of $-\rho_5$ is applied in all other cases, such as when the UL-SCH is idle (given to all UEs) or when a duplicate packet is decoded at the BS (given to the UE that transmitted the packet).

To train NPM using the reward function in \eqref{eq:reward}, we adopt a batch training method based on experience replay memory, which has been widely used to improve training stability \cite{Mnih:2013}. The experience tuple at training time slot $t'$ in episode $n$ is defined as $e_n^{t'} = \left(s_n^{t'}, a_n^{t'}, r_n^{t'}, s_n^{t'+1}\right)$, where the state, action, and reward are given by $s_n^{t'} = \bigcup_{\ell=0}^L b_{n,\ell}^{t'}, \text{ } a_n^{t'} = \bigcup_{\ell=1}^L a_{n,\ell}^{t'}, \text{ and } r_n^{t'} = \bigcup_{\ell=1}^L r_{n,\ell}^{t'}.$ The replay memory at training time slot $t'$ in episode $n$ is denoted by $\mathcal{D}_n^{t'}$, which stores all experience tuples until episodes $n-1$, i.e., $\bigcup_{\nu=1}^{n-1} \bigcup_{\tau=0}^{\lfloor T/B \rfloor - 1} e_\nu^{\tau}$, as well as those up to time $t'$ in episode $n$, i.e., $\bigcup_{\tau=0}^{t'} e_n^{\tau}$, where $\lfloor \cdot \rfloor$ is the floor function.


At each training time slot $t'$, a fixed-size batch of experience tuple $e_\nu^{\tau}$ is sampled from $\mathcal{D}_n^{t'}$ and used to train $\boldsymbol{\Theta}_n^{t'}$ by minimizing the average temporal-difference (TD) loss \cite{Sutton:1988}:
\begin{align}
    \mathcal{L}^{\text{TD}} \!\!=\!\! \frac{1}{L} \!\sum_{\ell=1}^L\! \mathbb{E}\!
    \Big[ \!\bigg\{ \!r_{\nu,\ell}^{\tau} \!+\! \gamma Q^{'}\!\!\left(\!s_{\nu,\ell}^{\tau+1}\!, a_{\nu,\ell}^{\tau+1}\!; \!\boldsymbol{\hat{\Theta}}_n^{t'}\!\right) 
    \! \nonumber  \! \\- \!Q\!\left(\!s_{\nu,\ell}^{\tau}, a_{\nu,\ell}^{\tau}; \!\boldsymbol{\Theta}_n^{t'}\!\right) \!\!\bigg\}^2 \Big],
    \label{eq:TD loss}
\end{align}
where the expectation is taken over $(s_{\nu,\ell}^{\tau}, a_{\nu,\ell}^{\tau},r_{\nu,\ell}^{\tau},s_{\nu,\ell}^{\tau+1}) \in e_\nu^{\tau}$, and $\gamma < 1$ is the discount factor. The functions $Q\left(\cdot; \boldsymbol{\Theta}\right)$ and $Q^{'}\left(\cdot; \boldsymbol{\hat{\Theta}}\right)$ represent the current and target Q-value networks, respectively. The target network shares the same architecture as the current network but uses frozen parameters that are updated less frequently to ensure training stability. The target network parameters are updated using a soft update mechanism \cite{Lillicrap:2015}:
$\boldsymbol{\hat{\Theta}}_n^{t'} \leftarrow \left(1 - \sigma\right)\boldsymbol{\hat{\Theta}}_n^{t'-1} + \sigma \boldsymbol{\Theta}_n^{t'},$ where $\sigma$ is a hyperparameter that controls the update speed. 
We consider a system where $\boldsymbol{\Theta}_n$ is updated every $T$ TTIs, such that $\boldsymbol{\Theta}_n^{\lfloor T/B \rfloor} = \boldsymbol{\Theta}_{n+1}$
for the testing phase of episode $n+1$. The same parameter set is also used for the initial training phase of episode $n+1$ as $\boldsymbol{\Theta}_n^{\lfloor T/B \rfloor} = \boldsymbol{\Theta}_{n+1}^{0}$.

\section{TPM: Token-based Protocol Model}
\label{section:TPM}
In NPM, when $L$ increases, the BS fails to construct DCMs since the input-output dimension of the function $f^D\left(\cdot;\boldsymbol{\theta}_n^D\right)$ varies, requiring re-training. To address this limitation, we propose a TPM that enables immediate action decisions without re-training, leveraging the flexibility of LLMs in handling variable-length input and output sentences, along with their ability to comprehend task instructions embedded in natural language. In TPM, increased $L$ merely lengthens the input prompt, which the LLM can process without any additional training or fine-tuning, as long as the prompt remains within its context window size.

The main problem of TPM is that its performance is sensitive to how the task is described in the instruction prompt \cite{Liu:2024_3}, necessitating prompt engineering. Prompt engineering involves designing task instructions either heuristically or automatically. In this paper, we adopt TextGrad \cite{Yuksekgonul:2024} to automatically generate the instruction. Once designed, the same instruction can be reused across different environments, as it describes the task itself rather than the environment-specific parameters.

\begin{figure}[t]
\centering
\includegraphics[width=0.93\columnwidth]{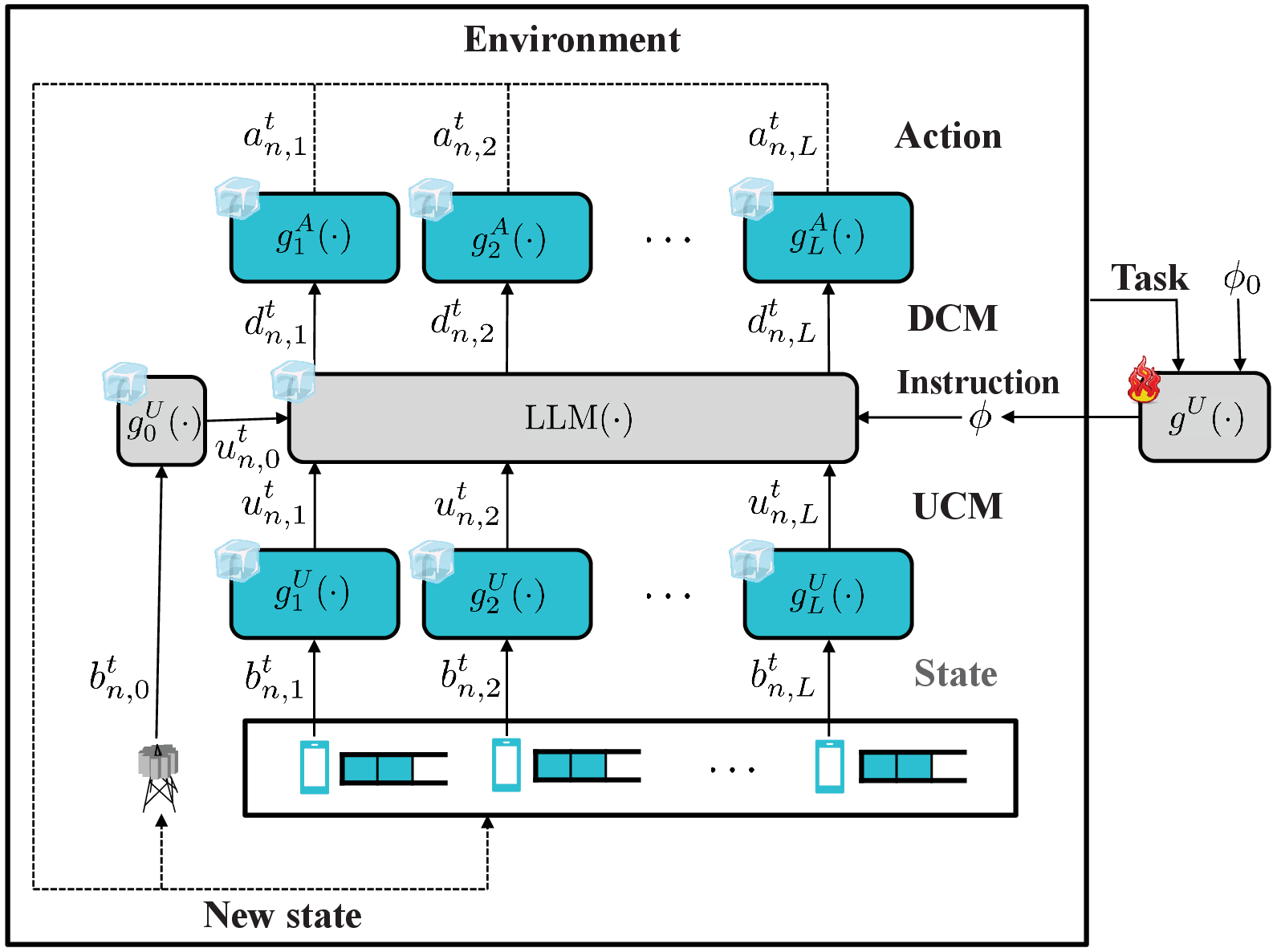}
\caption{A schematic illustration of TPM.}
\vspace{-1em} 
\label{fig:TPM}
\end{figure}

\subsection{TPM Construction and Operation}
\label{subsection: TPM construction and operation}
Unlike the uninterpretable control messages in NPM, the UCMs and DCMs in TPM are expressed in natural language, based on the architecture depicted in Fig.~\ref{fig:TPM}. 
Precisely, each UE $\ell$ constructs and transmits its UCM as $u_{n,\ell}^{t} = g_{\ell}^U\left(b_{n,\ell}^{t}\right),$ where $g_{\ell}^U\left(\cdot\right)$ is a function that converts UE $\ell$'s observation into a natural language query. The structure of the query is fixed, and the observation $b_{n,\ell}^{t}$ is inserted into a designated placeholder within this predefined format, as illustrated in the \textbf{\# Query} section of Fig.~\ref{fig:TPM example}. As the space for expression in the query is constrained, for consistency, we heuristically design the query format rather than automatically optimizing it.

The BS constructs and transmits the DCMs, the response generated by the LLM, to indicate the actions for all UEs as
\begin{align}
\bigcup_{\ell=1}^L d_{n,\ell}^{t} = \text{LLM}\left(\phi, \bigcup_{\ell=0}^L u_{n,\ell}^{t}\right),
\end{align}
where the function $\text{LLM}\left(\cdot\right)$ denotes the response of the LLM given a set of inputs. In addition to the UCMs received from all UEs, the LLM also requires an instruction prompt $\phi$ and a query representing the BS observation $u_{n,0}^{t}$. The BS query is generated as $u_{n,0}^{t} = g_{0}^U\left(b_{n,0}^{t}\right)$, where $g_{0}^U\left(\cdot\right)$ is a function that converts the BS observation into a natural language query. The instruction $\phi$ is predefined before TPM operation begins and is constructed using the prompt engineering function $g^U\left(\cdot\right)$, which describes the task. Unlike $\boldsymbol{\Theta}_n$ in NPM, which is updated in every episode to adapt to the task and environment through re-training, the LLM parameters remain fixed throughout all episodes.

After receiving the DCM $d_{n,\ell}^{t}$, each UE determines its action, which is explicitly indicated in the DCM, as $a_{n,\ell}^{t} = g_{\ell}^A\left(d_{n,\ell}^{t}\right),$ where $g_{\ell}^A\left(\cdot\right)$ is a function that converts the natural language response into a discrete action. Since the LLM output includes both natural language and the action tokens, a fixed response format for $d_{n,\ell}^{t}$ is necessary to enable correct parsing by $g_{\ell}^A$. For example, as shown in the \textbf{\# Answer} section of Fig.~\ref{fig:TPM example}, the fixed format ensures that the token `1' in the action field is not confused with `Agent 1:' in the natural language, by defining the expected position of the action token within the response. If the LLM generates a response that violates the fixed format due to hallucination, the action becomes unidentifiable. In such cases, to prevent potential collisions, the action is set as $a_{n,\ell}^{t} = 0$ (i.e., silent). To reduce the hallucinations, the instruction $\phi$ includes not only a task description but also a specification of the response format for UE actions. An example of such an instruction is shown in the \textbf{\# Instruction} section of Fig.~\ref{fig:TPM example}.

\begin{figure}[t]
\centering
\begin{ttcolorbox}\footnotesize
\textbf{$\mathbf{\#}$ Instruction} ($\phi$)\\
Base Station (BS): Controls user equipment (UE) communications.\\
UEs: Choose one of the following actions: Action 0 (wait), Action 1 (transmit), or Action 2 (delete).\\
Rules:\\
- Only one UE should transmit at a time to avoid collisions.\\
- UEs must delete packets that have already been successfully decoded by the BS.\\
- UEs should not delete packets that have not been decoded.\\
- Avoid transmitting or waiting on packets that have already been decoded, as this wastes time.\\
- Prevent collisions and packet loss by following these rules.\\
Provide answers in the format: `UE $\#$: Action $\#$'.\\

\textbf{$\mathbf{\#}$ Query} ($u_{n,1}^{t},u_{n,2}^{t}$,$u_{n,3}^{t}$,$u_{n,0}^{t}$)\\
UE 1 has 2 packets in the buffer.\\
UE 2 has 1 packet in the buffer.\\
UE 3 has 1 packet in the buffer.\\
BS successfully decoded Agent 2's packet. \\
Which action should each UE choose right now? \\

\textbf{$\mathbf{\#}$ Answer} ($d_{n,1}^{t},d_{n,2}^{t}$,$d_{n,3}^{t}$) \\
UE 1: Action 1\\
UE 2: Action 2 \\
UE 3: Action 0 
\end{ttcolorbox}
  \caption{TPM example.}
  \vspace{-1em} 
  \label{fig:TPM example}
\end{figure}

\subsection{TextGrad-aided TPM}
\label{subsection:TextGrad-TPM}
The goodput of TPM depends heavily on the LLM's ability to understand the context of the provided instruction and queries. However, since LLMs are typically trained for general purposes using large-scale datasets, they often lack domain-specific knowledge required for communication tasks \cite{Bariah:2023_2}. Therefore, domain adaptation is necessary to align the LLM with the specific requirements of the MAC protocol task. Three common approaches are used to tailor LLMs to specific tasks: fine-tuning, in-context learning (ICL), and prompt engineering \cite{Shin:2023}. Fine-tuning involves adjusting model parameters, which incurs substantial computational overhead and is not feasible in real time, making it unsuitable for resilient protocol operations. Without adjusting model parameters, ICL adapts the LLM by updating its context using exemplary few-shot queries. However, such queries are often unavailable immediately after environmental shifts. This makes ICL unsuitable as well, as we will elaborate further through simulation in Sec.~\ref{Sec:Ablation}.

Prompt engineering, by contrast, steers the LLM's responses by refining its input prompts, and thus remains applicable regardless of environmental shifts without requiring model parameter updates. It is particularly effective for refining instructions that remain valid even after environmental shifts.
While prior work in prompt engineering has focused largely on heuristic, handcrafted instruction design \cite{Liu:2024_3}, TextGrad \cite{Yuksekgonul:2024} enables automatic prompt refinement using an LLM. The instruction is iteratively refined by reflecting the produced feedback via the LLM, i.e., $g^U\left(\cdot\right) = \text{LLM}\left(\cdot\right)$. In the TextGrad-aided TPM, only the prompt generation function $g^U\left(\cdot\right)$, which designs the task instruction, is trainable.
This function offers the largest flexibility in expression, and even minor modifications to its output significantly influence TPM's performance.

\begin{figure}[t]
\centering
\begin{ttcolorbox} \footnotesize
$\#$  \textbf{Previous instruction} $ \left(\phi_{m}\right)$\\
Base Station (BS): Controls user equipment (UE) communications. \\
UEs: Choose one of these actions: Action 0 (wait), 1 (transmit), or 2 (delete). \\
Prevent collision, packet loss.
Please provide answers in the format `UE $\#$: Action $\#$'. \\

$\#$  \textbf{Feedback}
$\left(\frac{\partial \mathcal{L}^{\text{TG}}}{\partial \phi_{m}}\right)$\\
The system prompt should explicitly state the following key rules to align with the objective function: \\
1. Only one UE should transmit at a time to avoid collisions. \\
2. UEs must delete packets that have already been successfully decoded by the BS. \\
3. UEs should not delete packets that have not been decoded. \\
4. Avoid waiting or transmitting packets that have already been decoded, as this wastes time.\\
Adding these constraints will help the language model consistently optimize for successful, efficient packet transmissions while avoiding unnecessary actions.\\

$\#$  \textbf{Updated instruction} $\left(\phi_{m+1}\right)$\\
Updated instruction is \textbf{$\mathbf{\#}$ Instruction} ($\phi$) in Fig.~\ref{fig:TPM example}.
\end{ttcolorbox}
  \caption{TextGrad example at epoch $m$.}
  \vspace{-1em} 
  \label{fig:textgrad}
\end{figure}

In TextGrad, the processes of conventional NN training: feed-forward, backpropagation, and gradient descent, are mapped to response generation (textual feed-forward), feedback acquisition (textual backpropagation), and instruction update (textual gradient descent) at each training epoch $m$, starting from an initial hand-crafted instruction $\phi_0$. The textual feed-forward step generates a response $s_{m}$ using the current instruction $\phi_m$ and the query input $x$, as
\begin{align}
s_{m} = \text{LLM}\left(\phi_{m}, x\right). \label{eq:textual feedforward}
\end{align}
The training data $x$ is acquired by $x = \bigcup_{\ell=0}^L g_{\ell}^U\left( b_{\ell}\right),$ where $b_{\ell}$ is a hand-crafted observation value used for TextGrad training. If the current instruction $\phi_m$ is suboptimal, feedback is generated in the textual backpropagation step as
\begin{align}
\frac{\partial \mathcal{L}^{\text{TG}}}{\partial \phi_{m}} = \text{LLM}\left(\phi_{m}, x, s_{m}, \mathcal{L}^{\text{TG}}\right),
\label{eq:textual backpropagation}
\end{align}
where $\mathcal{L}^{\text{TG}}$ is the textual objective function, defined as $\mathcal{L}^{\text{TG}} = g_{r}^U\left(C\left(r_{n,\ell}^{t'}\right)\right),$
where $C\left(\cdot\right)$ produces all the conditions required to earn a different reward $r_{n,\ell}^{t'}$, and $g_r^U\left(\cdot\right)$ converts these conditions into natural language. Unlike gradients in conventional NN training, the feedback in TextGrad is in natural language, providing interpretability for both humans and the LLM itself. Note that the partial derivative here is not a numerical gradient, but a symbolic representation of the feedback process. The instruction is then updated during the textual gradient descent step using this feedback as
\begin{align}
\phi_{m+1} = \text{LLM}\left(\phi_{m}, \frac{\partial \mathcal{L}^{\text{TG}}}{\partial \phi_{m}}\right). \label{eq:textual gradient descent}
\end{align}
An example of instruction update at epoch $m$ is illustrated in Fig.~\ref{fig:textgrad}, in which the updated instruction is exploited in Fig.~\ref{fig:TPM example}.

The optimization process terminates when the feedback indicates that the instruction aligns with the objective function, i.e., when $\partial \mathcal{L}^{\text{TG}} / \partial \phi_{m} \approx 0$, or when a predefined maximum number of epochs is reached. The final instruction $\phi$ is selected to be the instruction $\phi_{m}$ that achieves the highest goodput across TPM operations as $\phi = \arg\max_{\phi_m} G^{\phi_m}$, where $G^{\phi_m}$ is the goodput obtained by instruction $\phi_m$. The episode index $n$ is omitted since TPM’s LLM parameter set remains constant across episodes. Once $\phi$ is optimized, it is applied consistently across all environments, as it includes only the task description and is independent of specific environment parameters.

\section{T2NPM: TPM-to-NPM}
\label{section:T2NPM}
TPM enables immediate action decisions after an environmental shift without requiring re-training. However, despite the benefits of automatic instruction design, TPM remains only partially grounded in the environment, as it is not fine-tuned to a task-specific environment. Consequently, if the environment remains stationary after the environmental shift, TPM may select suboptimal actions compared to a re-trained NPM. To address this limitation, we introduce T2NPM, a method that accelerates the re-training of NPM using KD under a stationary environment. This approach leverages KD's advantage that can significantly accelerate NN training \cite{Phuong:2019}. In T2NPM, the student model (NPM) is trained with guidance from the teacher model, the TextGrad-aided TPM.

\subsection{Teacher Knowledge Construction}
In the original TextGrad-aided TPM, the function $g_\ell^A\left(\cdot\right)$ returns only the selected action $a_{n,\ell}^{t}$. However, KD requires that the teacher and student models produce outputs within the same domain. Specifically, both models must output logits corresponding to all possible actions in $\mathcal{A}$. To satisfy this requirement, we modify the teacher model’s output to include logits only over $\mathcal{A}$ rather than the token vocabulary space $\mathcal{T}$, using the result that token generation probabilities in LLMs are correlated with prediction correctness \cite{Plaut:2024}.

From the LLM output, the token generation probabilities for the action tokens `1,' `2,' and `0' (as shown in the \textbf{\# Answer} of Fig.~\ref{fig:TPM example}) are extracted from $\mathcal{T}$ and then restricted to $\mathcal{A}$. The LLM output $d_{n,\ell}^{t'}$ contains the token $y_{n,\ell}^{t'}$ that determines the action for UE $\ell$. If $d_{n,\ell}^{t'}$ follows the answer format defined by the instruction $\phi$, this action token $y_{n,\ell}^{t'}$ is identifiable. Let LLM response $d_{n,\ell}^{t'}$ be a sequence of tokens $d_{n,\ell}^{t'} = \left(y_1, y_2, \ldots, y_J\right),$ where $y_j$ denotes the $j$-th token, and $J$ is the total number of tokens in the response. Among these, let $j^*$-th token correspond to the action indicator, i.e., $y_{n,\ell}^{t'} = y_{j^*}$. For instance, $y_{n,1}^{t'} = y_{j^*} = \text{`2'},$ if the response is `UE 1 action: 2.' The token generation probability of the entire response given the instruction and UCMs $\mathrm{Pr}\left(d_{n,\ell}^{t'} \mid \phi, \bigcup_{\ell=0}^L u_{n,\ell}^{t'}\right)\in \mathbb{R}^{|\mathcal{T}|}$, is expressed as
\begin{align}
\hspace{-5pt}\mathrm{Pr}\!\left(d_{n,\ell}^{t'} \mid \phi, \bigcup_{\ell=0}^L u_{n,\ell}^{t'}\right) 
= \prod_{j=1}^{J} \mathrm{Pr}\!\left(y_j \mid \phi, \bigcup_{\ell=0}^L u_{n,\ell}^{t'}, y_{<j}\right),
\end{align}
where $y_{<j}$ denotes all tokens preceding position $j$. Focusing at position $j^*$, the LLM computes a conditional distribution over $\mathcal{T}$ as $\mathrm{Pr}\left(y_{j^*} \mid \phi, \bigcup_{\ell=0}^L u_{n,\ell}^{t'}, y_{<j^*}\right) \in \mathbb{R}^{|\mathcal{T}|}$. To align with the output space of the student model, the dimension of this distribution is restricted to $\mathbb{R}^{|\mathcal{A}|}$.

To extract the action probabilities at $j^*$, the logits corresponding to tokens in $\mathcal{A}$ are isolated. A masked softmax is then applied over $\mathcal{A}$ as 
\begin{align}
&\mathrm{Pr}\bigg(y_{j^*} = \text{`$a$'} \mid \phi, \bigcup_{\ell=0}^L u_{n,\ell}^{t'}, y_{<j^*}\bigg) \nonumber\\
&\hspace{-10pt}= \frac{\exp\left(\mathrm{Pr}\!\left(y_{j^*} = \text{`$a$'} \mid \phi, \bigcup_{\ell=0}^L u_{n,\ell}^{t'}, y_{<j^*}\right) / \kappa \right)}
{\sum\limits_{w \in \mathcal{A}} \exp\left(\mathrm{Pr}\!\left(y_{j^*} = \text{`$w$'} \mid \phi, \bigcup_{\ell=0}^L u_{n,\ell}^{t'}, y_{<j^*}\right) / \kappa \right)},
\label{eq:softmax}
\end{align}
where $\kappa$ is the temperature parameter, and $\text{`$a$'}, \text{`$w$'} \in \{\text{`0',`1',`2'}\}$ is a character representation of an integer element in $\mathcal{A}$. A higher temperature $\kappa$ produces a softer probability distribution, which helps retain more information about the relative information across classes, at the expense of reduced prediction confidence. For simplicity, we denote $\mathrm{Pr}\left(y_{j^*} = \text{`$a$'} \mid \phi, \bigcup_{\ell=0}^L u_{n,\ell}^{t'}, y_{<j^*}\right)$ as $\mathrm{Pr}\left(y_{n,\ell}^{t'} = \text{`$a$'}\right)$. This masked softmax ensures that the output probabilities are normalized exclusively over $\mathcal{A}$, thereby eliminating the influence of irrelevant tokens from $\mathcal{T}$.

The teacher logits for UE $\ell$, denoted by $\mu_{n,\ell}^{t'}$, are extracted by concatenating the token generation probabilities corresponding to all actions in $\mathcal{A}$. Specifically, $\mu_{n,\ell}^{t'} = \left[\mathrm{Pr}\left(y_{n,\ell}^{t'} = \text{`$0$'}\right),\, \mathrm{Pr}\left(y_{n,\ell}^{t'} = \text{`$1$'}\right),\, \mathrm{Pr}\left(y_{n,\ell}^{t'} = \text{`$2$'}\right)\right].$
If the action token $y_{n,\ell}^{t'}$ is unidentifiable, i.e., when $d_{n,\ell}^{t'}$ deviates from the answer format specified in $\phi$, we default to a uniform distribution as
$\mu_{n,\ell}^{t'} = [1/3,\, 1/3,\, 1/3].$
The aggregated teacher knowledge from all $L$ UEs is represented by $M_{n}^{t'} = \bigcup_{\ell=1}^L \mu_{n,\ell}^{t'}.$
It is important to note that the teacher model in T2NPM is primarily used to guide and regularize the early stages of student training, rather than to transfer optimal action policies. As such, the teacher model is not required to generate optimal decisions. Self-distillation \cite{Furlanello:2018}, where the same NN can act as both teacher and student, is a supporting example since it still benefits from KD, even when the teacher model is not optimal.

\subsection{T2NPM Construction via Knowledge Distillation}
T2NPM is trained using a composite loss function that combines the TD loss defined in \eqref{eq:TD loss} with a KD loss $\mathcal{L}^{\text{KD}}$. To compute the KD loss, we introduce a teacher replay memory $\tilde{\mathcal{D}}_n^{t'} = \left\{ s_\nu^{\tau} \mid s_\nu^{\tau} \in \bigcup_{e_\nu^{\tau} \in \mathcal{D}_n^{t'}} e_\nu^{\tau},\; 1 \leq \nu \leq n,\; 0 \leq \tau \leq t' \right\},$ which stores only the states of experience tuples from $\mathcal{D}_n^{t'}$. At each student update step, a training batch of states is sampled from $\tilde{\mathcal{D}}_n^{t'}$, denoted as $\tilde{s}_n^{t'}$. For each sampled state $\tilde{s}_n^{t'}$, UE $\ell$ generates a corresponding DCM, $\tilde{d}_{n,\ell}^{t'}$. The student model then computes its logits using $\tilde{\pi}_{n,\ell}^{t'} = f^A\left(\tilde{d}_{n,\ell}^{t'}; \boldsymbol{\theta}_{n,\ell}^{A,t'}\right) \in \mathbb{R}^{|\mathcal{A}|},$ which represents the Q-values across all actions in $\mathcal{A}$. To obtain soft logits, a temperature-scaled softmax is applied as $\mathrm{Pr}\left(a_{n,\ell}^{t'} = w\right) = \frac{\exp\left(\tilde{\pi}_{n,\ell}^{t'}[w]/\kappa\right)}{\sum_{w'=0}^{2} \exp\left(\tilde{\pi}_{n,\ell}^{t'}[w']/\kappa\right)},$ where $\kappa$ is the temperature parameter, and 
$\tilde{\pi}_{n,\ell}^{t'}[w]$ denotes the logit corresponding to a class $w \in \mathcal{A}$. The resulting soft logits for UE $\ell$ are then concatenated into $\pi_{n,\ell}^{t'} = \left[ \mathrm{Pr}\left(a_{n,\ell}^{t'} = 0\right),\, \mathrm{Pr}\left(a_{n,\ell}^{t'} = 1\right),\, \mathrm{Pr}\left(a_{n,\ell}^{t'} = 2\right) \right].$ The aggregated student knowledge across all UEs is denoted as $\Pi_{n}^{t'} = \bigcup_{\ell=1}^{L} \pi_{n,\ell}^{t'}.$ Since both the teacher’s and student’s knowledge, $M_n^{t'}$ and $\Pi_n^{t'}$, are constructed using the same sampled state $\tilde{s}_n^{t'}$, a direct comparison of soft logits is enabled during training.

The KD loss is computed using the Kullback–Leibler divergence (KLD) between the teacher knowledge $M_n^{t'}$ and the student knowledge $\Pi_n^{t'}$, as
\begin{align}
    \mathcal{L}^{\text{KD}} = \mathbb{E}_{s_n^{t'} \sim \tilde{\mathcal{D}}_n^{t'}}\left[D_{\mathrm{KL}}\left(M_n^{t'} \,\Vert\, \Pi_n^{t'}\right)\right],
    \label{eq:KDloss}
\end{align}
where the KLD is calculated by
\begin{align}
    &D_{\mathrm{KL}}\left(M_n^{t'} \,\Vert\, \Pi_n^{t'}\right) \nonumber\\&= 
    \sum_{\ell=1}^{L} \sum_{w=0}^{2} \mathrm{Pr}\left(y_{n,\ell}^{t'} = \text{`$w$'}\right) 
    \log \frac{\mathrm{Pr}\left(y_{n,\ell}^{t'} = \text{`$w$'}\right)}{\mathrm{Pr}\left(a_{n,\ell}^{t'} = w\right)}.
\end{align}
The final loss function used to train T2NPM is a weighted combination of the TD loss and KD loss
\begin{align}
    \mathcal{L} = \lambda_1 \mathcal{L}^{\text{TD}} + \lambda_2 \mathcal{L}^{\text{KD}},
    \label{eq:T2NPM loss function}
\end{align}
where $\lambda_1, \lambda_2 > 0$ are hyperparameters that balance the contributions of the TD loss and the KD loss. The proposed regularization term $\mathcal{L}^{\text{KD}}$ accelerates convergence in the early stages of NN training, as will be demonstrated in the simulation results in Sec.~\ref{section:simulation}.

\section{T3NPM: TPM-after-T2NPM}
\label{section:T3NPM}
TPM enables immediate action decisions under environmental shifts, whereas the re-trained T2NPM demonstrates superior long-term performance. To leverage the strengths of both approaches, we propose T3NPM, a hybrid protocol model that employs TPM during the early stage of T2NPM re-training and transitions to T2NPM once sufficient training has been conducted. The transition timing from TPM to T2NPM is determined via hypothesis testing based on multiple goodput measurements from the ongoing T2NPM re-training, which reduces the available re-training time due to the fixed single episode duration and the separation between the re-training and measurement phases. We exploit the proposed novel meta-resilience metric to effectively balance this trade-off, where longer measurement allows for more reliable comparison but shortens re-training time, which is crucial for goodput improvement. Unlike conventional resilience metrics designed for environments with known target performance \cite{Najarian:2019}, meta-resilience quantifies the system's resilience under environmental shifts, where the target performance is unknown. As we elaborate in the following subsections, this metric enables both fair performance comparisons across various protocol models and optimization of T3NPM’s transition timing.

\begin{figure}[t]
    \begin{subfigure}{0.45\textwidth}
    \centering
    \includegraphics[width=0.9\textwidth]{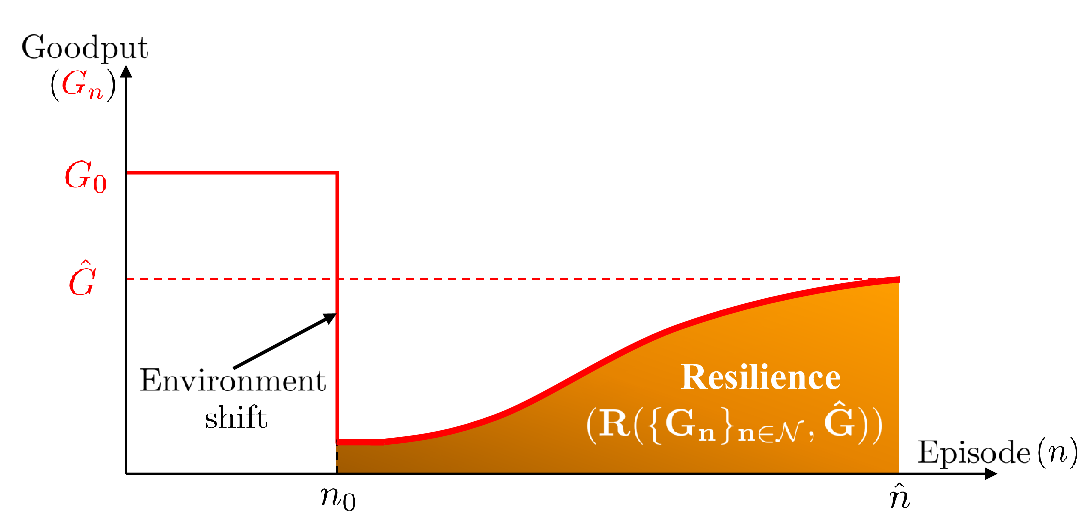}
    \caption{\centering Resilience (AUC of goodput).}
\label{fig:resilience}
    \end{subfigure}
    \vspace{0.2em}
    \begin{subfigure}{0.45\textwidth}
    \centering
    \includegraphics[width=0.9\textwidth]{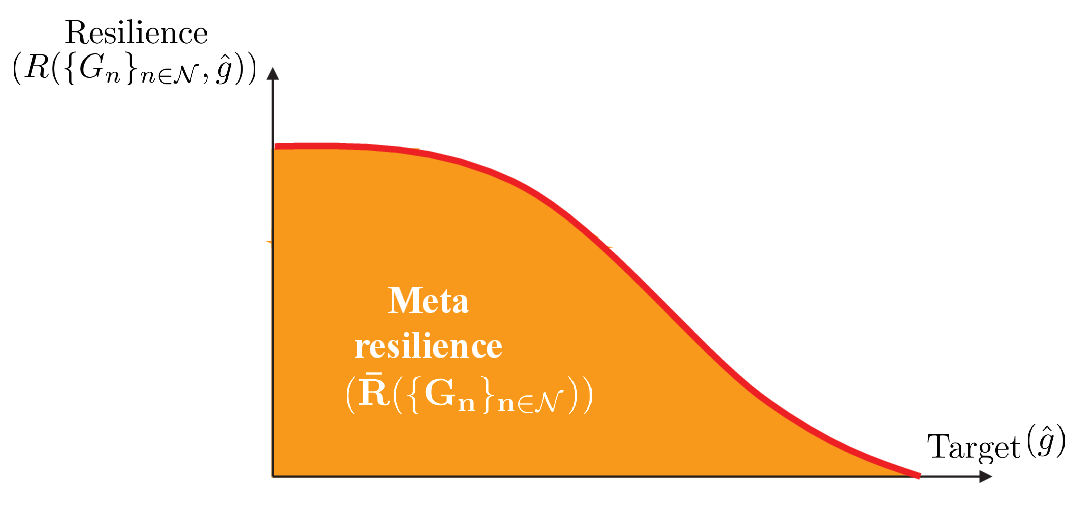}
    \caption{\centering Meta-resilience (AUC of resilience).}
\label{fig:meta-resilience}
    \end{subfigure}
    \caption{Resilience vs. meta-resilience.}
    \vspace{-1em} 
\label{fig:resilience vs meta-resilience}
\end{figure}

\subsection{Meta-resilience}
\label{subsection:Meta-resilience}
Since our proposed methods focus on environmental shifts, resilience is used as an evaluation metric. In \cite{Cetinkaya:2013}, resilience is defined as the ability of a network to maintain acceptable service levels, reflecting a similar perspective of \cite{Reed:2009}. Quantitative definition is proposed at \cite{Reifert:2023}, which is the combined metrics of resilience and mixed criticality, building on the framework of \cite{Najarian:2019}. Among the proposed definitions in \cite{Najarian:2019}, the metric adapted from \cite{Bruneau:2003}, which uses the area under the curve (AUC), is particularly suited for wireless communication systems due to its simplicity and general applicability. As illustrated in Fig.~\ref{fig:resilience}, we define resilience as
\begin{align}
R\left(\{G_n\}_{n\in \mathcal{N}}, \hat{G}\right) = \frac{1}{|\mathcal{N}|}\sum_{n \in \mathcal{N}} \min\left\{ \frac{G_n}{\hat{G}},\, 1 \right\} ,
\label{eq:resilience}
\end{align}
where $G_n$ denotes the system goodput at episode $n$, and $\hat{G}$ is a constant representing the target goodput. This normalization reflects the ratio between the AUC of a constant-performance rectangle at height $\hat{G}$ and the actual AUC of the observed performance trajectory. The $\min\{\cdot\}$ operator ensures that instances where $G_n > \hat{G}$ do not dominantly affect the metric, thereby preventing divergence.

If the environmental shift does not occur, $\hat{G}$ in \eqref{eq:resilience} can be set as $G_0$, representing the pre-environmental shift performance baseline, as illustrated in Fig.~\ref{fig:resilience}. This corresponds to a typical disaster-recovery scenario where the system attempts to return to a known target performance level. However, a key limitation of \eqref{eq:resilience} arises when $\hat{G}$ is undefined, which is often the case under environmental shifts, especially when the post-shift performance is unpredictable. To address this, we propose a novel meta-resilience metric that evaluates system resilience in the absence of a fixed target performance. The term `meta' reflects a holistic view of system performance across a range of operating conditions, analogous to the use in other research domains. Notable examples include meta-learning \cite{Finn:2017}, which leverages knowledge from multiple tasks to enable rapid adaptation to new tasks, and meta-distribution, which captures variability or uncertainty across probability distributions \cite{Haenggi:2015}.

\begin{figure}[t]
\centering
\includegraphics[width=\columnwidth]{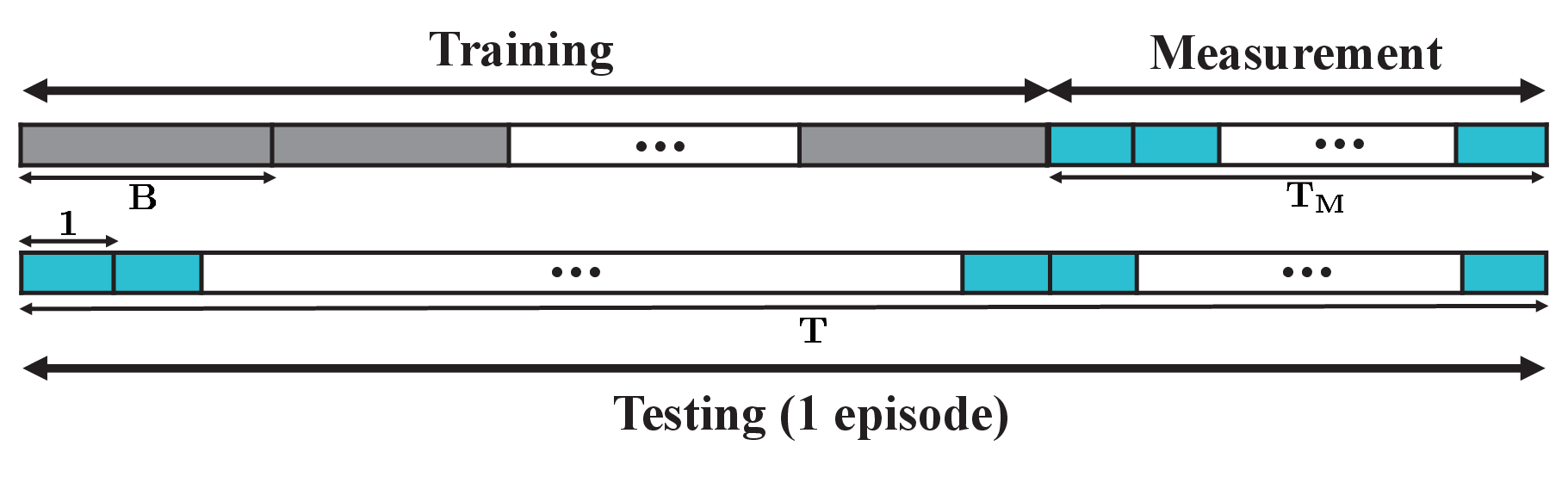}
\caption{Episode temporal structure, comprising measurement time $T_M$ and training time $T-T_M$.}
\vspace{-1em} 
\label{fig:Single episode}
\end{figure}

In the context of this work, we focus on the scenario of an increasing number of UEs. While this increase incurs significant environmental shift from the perspective of NN, it is a common situation in practical systems, rather than a malfunction requiring repair. A fixed $\hat{G}$ cannot be defined in such cases, as the oracle policy in the shifted environment is unknown. To address this, we introduce the concept of meta-resilience $\bar{R}\left(\{G_n\}_{n\in \mathcal{N}}\right)$, which evaluates the resilience of the system across a predetermined feasible range of target performances $\hat{g}$ instead of a fixed $\hat{G}$, defined as \begin{align}
\bar{R}\left(\{G_n\}_{n\in \mathcal{N}}\right) 
&= \frac{1}{|\mathcal{N}|}\int \sum_{n \in \mathcal{N}} \min\left(\frac{G_n}{\hat{g}},\, 1\right)   d\hat{g} \nonumber \\
&= \mathbb{E}_{\hat{g}} \left[ R\left(\{G_n\}_{n\in \mathcal{N}}, \hat{g}\right) \right].
\label{eq:meta resilience}
\end{align}
In practice, $\hat{g}$ represents possible levels of Quality-of-Service such as goodput, bit errors, or latency expected under varying network environments. The relationship between resilience and meta-resilience is illustrated in Fig.~\ref{fig:meta-resilience}. By integrating over a range of possible $\hat{g}$ values, meta-resilience offers a holistic measure of system adaptability under uncertain performance targets. In this paper, we use multiple goodput levels as realizations of $\hat{g}$, and hereafter treat $\hat{g}$ specifically as goodput.


\subsection{T3NPM Operation}
T3NPM aims to maximize meta-resilience by initially leveraging TPM's coarse-yet-immediate responses to environmental shifts, and then switching to T2NPM for its fine-tuned, environment-specific operations. The main challenge lies in determining when to switch. To support this decision-making, we employ the one-sided Mann-Whitney U test \cite{Reimers:2018}, a non-parametric statistical hypothesis test that does not assume Gaussian distributions, making it well-suited for NN-based hypothesis testing.

Precisely, TPM initially operates from episode $n_0$.
For episodes $n>n_0$, we define null and alternative hypotheses~as:
\begin{align}
H_{0}: & \quad \mathrm{median}\left(\mathcal{V}^{\text{TPM}}\right) \,\ge\,
         \mathrm{median}\left(\mathcal{V}_{n}^{\text{T2NPM}}\right), \label{eq:H0} \\[6pt]
H_{A}: & \quad \mathrm{median}\left(\mathcal{V}^{\text{TPM}}\right) \,<\,
         \mathrm{median}\left(\mathcal{V}_{n}^{\text{T2NPM}}\right), \label{eq:Ha}
\end{align}
where $\mathcal{V}^{\text{TPM}}$ and $\mathcal{V}_n^{\text{T2NPM}}$ denote the TPM and T2NPM goodput measurements at episode $n$, respectively. Given a threshold $\alpha>0$, if the U test returns a $p$-value, following the procedure in \cite{Nachar:2008},
, such that $p \geq \alpha$, the null hypothesis $H_0$ is accepted, and the protocol continues with TPM. If $p < \alpha$, $H_0$ is rejected, and the protocol switches to T2NPM that statistically outperforms TPM; specifically, $\mathrm{Pr}\left(v_1 > v_2\right) > 0.5$ for $v_1 \in \mathcal{V}_n^{\text{T2NPM}}$ and $v_2 \in \mathcal{V}^{\text{TPM}}$. As the goodput of T2NPM is statistically monotonically increasing with $n$, the switching decision remains fixed once made.

\begin{algorithm}[t]
\caption{T3NPM}
\label{alg:T3NPM}
\small
\textbf{Input:} Episode index set $\mathcal{N}$, MixSwitch parameters $T_{M}$, $k$.\\
\textbf{Output:} Goodput set $\{G_n\}_{n\in \mathcal{N}}$.
\begin{algorithmic}[1]
\State Initialize $\text{T2NPM}_{\text{Ind}} \gets 0$.
\State Initialize model parameters $\boldsymbol{\Theta}_{0}$.
\For{$n \in \mathcal{N}$} \Comment{Run training and testing phases in parallel}
    \State \textbf{Training Phase:}
    \State Train $\boldsymbol{\Theta}_n$ using \eqref{eq:T2NPM loss function}.
    \If{$n - k \ge n_0$ \textbf{and} $\text{T2NPM}_{\text{Ind}} = 0$}
        \State Generate $\tilde{\mathcal{V}}_{n}^{\text{T2NPM}}$ through $\boldsymbol{\Theta}_n$ within $T_M$ TTIs.
        \State Construct $\mathcal{V}_n^{\text{T2NPM}} = \bigcup_{k'=0}^{k} \tilde{\mathcal{V}}_{n-k'}^{\text{T2NPM}}$.
        \State Perform Mann-Whitney U test by setting null and alternative hypotheses as equations~\eqref{eq:H0} and~\eqref{eq:Ha}.
        \If{\eqref{eq:Ha} is accepted}
            \State $\text{T2NPM}_{\text{Ind}} \gets 1$.
        \EndIf
    \EndIf
    \State \textbf{Testing Phase:}
    \If{$n = n_0$ \textbf{or} $\text{T2NPM}_{\text{Ind}} = 0$}
        \State Operate TPM as described in Section~\ref{subsection: TPM construction and operation}.
    \Else
        \State Operate T2NPM as described in Section~\ref{subsection:NPM operation}.
    \EndIf
    \State Calculate $G_n$ using equation~\eqref{eq:goodput}.
\EndFor
\end{algorithmic}
\end{algorithm}

In this statistical hypothesis testing, decision accuracy improves with the number of goodput measurements, i.e., $|\mathcal{V}^{\text{TPM}}|$ and $|\mathcal{V}_n^{\text{T2NPM}}|$. Since TPM goodput remains statistically constant, $\mathcal{V}^{\text{TPM}}$ can be pre-computed, yielding a fixed $\phi = \arg\max_{\phi_m} G^{\phi_m}$ in Sec.~\ref{subsection:TextGrad-TPM} for all episodes. In contrast, T2NPM goodput evolves over re-training episodes, requiring measurement computation $\mathcal{V}_n^{\text{T2NPM}}$ at each episode. This leads to a time allocation trade-off between measurement and training phases. As illustrated in Fig.~\ref{fig:Single episode}, given the total $T$ TTIs per episode, allocating more $T_M$ TTIs to measurement improves switching decision accuracy but reduced the remaining training time to $T-T_M$ TTIs, potentially degrading T2NPM goodput. 

To alleviate this measurement-training trade-off, we propose MixSwitch, which reuses a fraction of past measurements. To reduce $T_M$ without significantly compromising T2NPM goodput, MixSwitch incorporates measurements from the past $k = \lceil \hat{T}_M / T_M \rceil - 1$ episodes, where $\hat{T}_M$ is the total TTIs required to construct $\mathcal{V}_n^{\text{T2NPM}}$ and $\lceil \cdot \rceil$ is the ceiling function. Accordingly, the measurement set is given by $\mathcal{V}_n^{\text{T2NPM}} = \bigcup_{k'=0}^{k} \tilde{\mathcal{V}}_{n-k'}^{\text{T2NPM}}$, where $k'\in [0,k]$ is the index over a window of size $k$, counting backwards from $n$, and $\tilde{\mathcal{V}}_n^{\text{T2NPM}}$ denotes the actual measurements collected during $T_M$ at episode $n$. The full operation of T3NPM with MixSwitch is summarized in Algorithm~1, with $\text{T2NPM}_{\text{Ind}}=1$ marking the switching event from TPM to T2NPM.
Given a fixed $k$, we aim to find an optimal $T_M$ that maximizes meta-resilience through the following optimization problem:
\vspace{-5pt}
\begin{equation}
\begin{array}{ll}
\displaystyle \max_{T_M} & \bar{R}\left(\{G_n\}_{n \in \mathcal{N}} \mid 
T_M, k\right) \\
\text{subject to} & 0 \le T_M \le T, \\
                  & k = \left\lceil \frac{\hat{T}_M}{T_M} \right\rceil - 1.
\end{array}
\label{p1:T3NPM opt}
\end{equation}
When  $T_M = 0$ (i.e., no measurement) or $T_M=T$ (i.e., no re-training), T3NPM becomes equivalent to T2NPM or TPM, respectively.
A heuristic solution $\hat{T}_M$ to \eqref{p1:T3NPM opt} and its effectiveness in improving meta-resilience will be presented in Sec.~\ref{section:simulation}.



\section{Numerical Results}
\label{section:simulation}
\subsection{Simulation Settings}
We consider a simulation scenario where training begins at $n_0 = 0$, and $L$ increases from $2$ to $3$, representing an environmental shift. The remaining simulation parameters are set as $p_\ell^a = 0.3$, $b_\ell^{\max} = 3$, and $p_{\ell}^e = 0.01$ for all $\ell$. Upon detecting the increase in $L$ at episode $n_0$, all UEs and the BS initiate re-training, which continues until $\hat{n} = 150$. The model update period during the testing phase is $T = 144$ TTIs, and the time interval between training updates, i.e., between $t'$ and $t'+1$, is $B = 4$ TTIs.  To stabilize the goodput calculation at the testing phase, we define a single goodput value within a single episode by averaging the number of successfully decoded packets over every $12$ TTIs. This process is repeated $12$ times within the episode, and the final reported goodput is obtained by averaging these $12$ goodput values.
During NPM training, $\epsilon$ is initialized at $1$ and decayed exponentially at a rate of $0.9$ per episode, with a lower bound of $0.1$. The reward values are configured as $\rho_1 = 10$, $\rho_2 = 8$, $\rho_3 = \rho_4 = 4$, and $\rho_5 = 1$. The discount factor and the soft-update parameter for the target network are set to $\gamma = 0.99$ and $\sigma = 10^{-3}$, respectively. For TPM, T2NPM, and T3NPM, we utilize the lightweight open-source SOLAR-10.7B LLM \cite{Kim:2023}. In T2NPM, $\kappa = 2$ to compare soft logits between TPM and NPM, and the composite loss function uses weights $\lambda_1 = 0.1$ and $\lambda_2 = 0.9$ for the TD and KD losses, respectively. In T3NPM, $k = 5$ for MixSwitch, and hypothesis testing is performed using the one-sided Mann–Whitney U test with $\alpha = 0.05$. As a classical benchmark, S-ALOHA is included with a fixed transmission probability of~$0.33$.

\begin{table}[t]
\captionsetup{justification=centering, labelsep=newline, font={smaller,sc}}
\caption{Goodput comparison under environmental shifts.}
\centering
\resizebox{\columnwidth}{!}{
\begin{tabular}{l|c|c|c|c|c|c|c|c}
\toprule
  &  \multicolumn{8}{c}{\textbf{Environmental Shift}} \\ 
\textbf{Scheme} 
& \textcolor{red}{$p_{\ell}^a \uparrow$}
& \textcolor{blue}{$p_{\ell}^a \downarrow$ }
& \textcolor{red}{$b_{\ell}^{\max} \uparrow$}
& \textcolor{blue}{$b_{\ell}^{\max} \downarrow$}
& \textcolor{red}{$p_{\ell}^e \uparrow$ }
& \textcolor{blue}{$p_{\ell}^e \downarrow$ }
& \textcolor{red}{$L \uparrow$}
& \textcolor{blue}{$L \downarrow$}
\\
\hline
\textcolor{gray}{S-ALOHA} & $\mathcolor{gray}{0.46}$ & $\mathcolor{gray}{0.28}$ & $\mathcolor{gray}{0.37}$ & $\mathcolor{gray}{0.37}$ &
$\mathcolor{gray}{0.33}$ &
$\mathcolor{gray}{0.37}$ &
$\mathcolor{gray}{0.37}$ &
$\mathcolor{gray}{0.30}$ \\
\textcircled{\raisebox{-0.85pt}{1}} NPM & $ 0.65 $ & $ 0.32 $ & $ 0.43 $ & $ 0.45 $ & $0.41$ & $0.45$ & $0.08$ & $0.06$\\
\textcircled{\raisebox{-0.85pt}{2}} Re-trained NPM\! & $ 0.76 $ & $ 0.42 $ & $ 0.58 $ & $ 0.55 $ & $0.54$ & $0.58$ & $0.51$ & $0.09$\\
\hline \hline
\textcircled{\raisebox{-0.85pt}{2}} - \textcircled{\raisebox{-0.85pt}{1}} & $ 0.11 $ & $ 0.10 $ & $ 0.15 $ & $ 0.10 $ & $0.13$ & $0.13$ & $\mathbf{0.43}$ & $0.03$\\
\bottomrule
\end{tabular}}
\label{tab: L increment validation}
\end{table}

\begin{figure}[t]
\centering
    \begin{subfigure}{0.84\columnwidth}
    \centering
    \includegraphics[width=0.83\columnwidth]{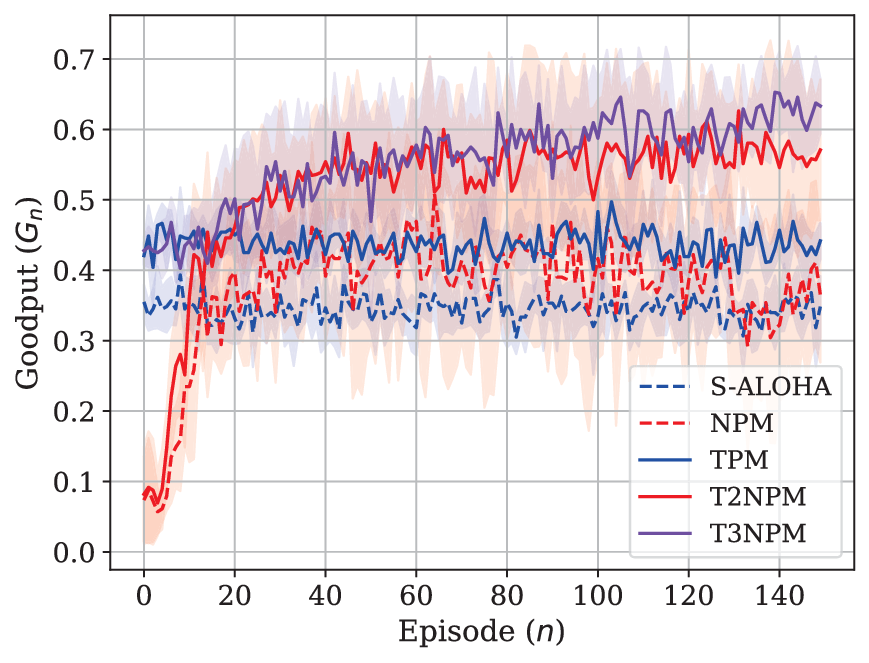}
    \caption{\centering Goodput during re-training under $L\uparrow$.}
    \label{sim_fig:R triangle methods}
    \end{subfigure}
    \vspace{0.1em}
    \begin{subfigure}{0.84\columnwidth}
    \centering
     \includegraphics[width=0.83\columnwidth]{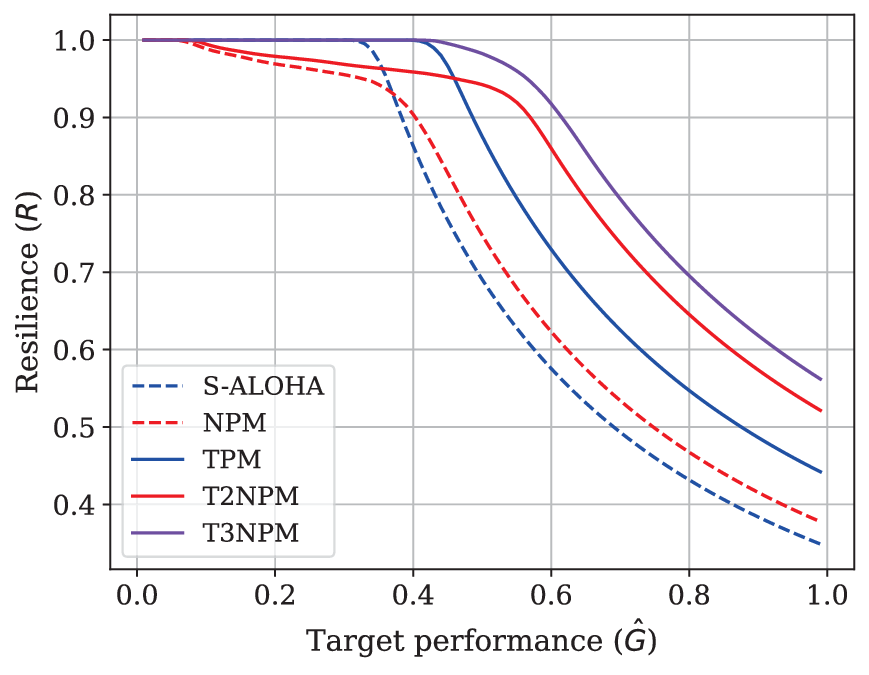}
    \caption{\centering Resilience w.r.t. a target goodput level $\hat{G}$.}
    \label{sim_fig:R methods}
    \end{subfigure}
    \vspace{0.1em}
    \begin{subfigure}{0.84\columnwidth}
    \centering
     \includegraphics[width=0.83\columnwidth]{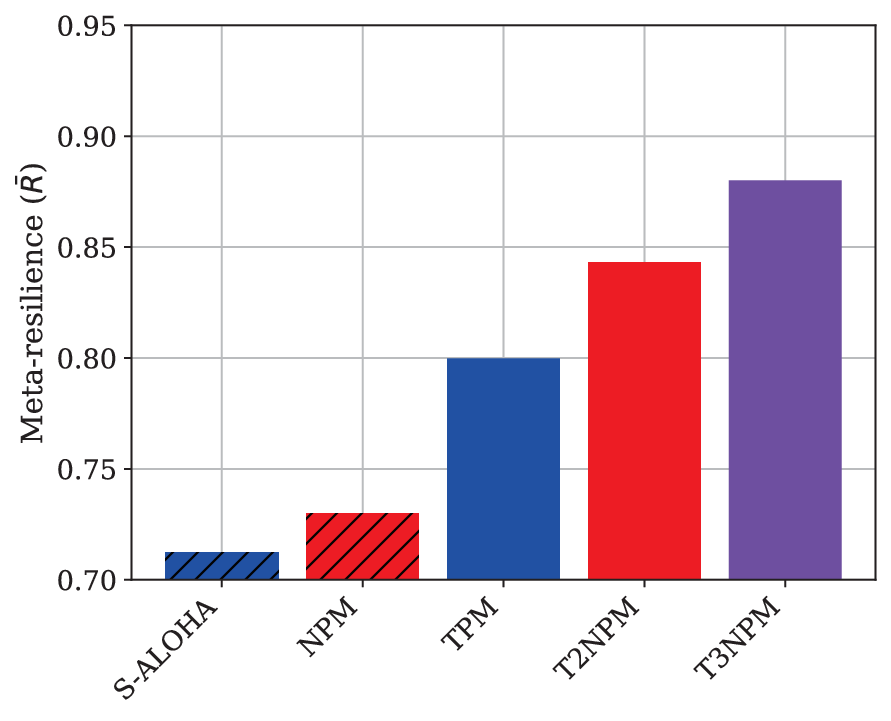}
    \caption{\centering Meta-resilience comparison.}
    \label{sim_fig:Meta-R methods}
    \end{subfigure}
\caption{\centering Goodput, resilience, and meta-resilience of semantic MAC protocols under the environmental shift $L\uparrow$.}
\vspace{-1.0em} 
\label{fig:numUE vs methods}
\end{figure}

\begin{figure*}[ht]
\centering
    \begin{subfigure}{0.32\textwidth}
    \centering
    \includegraphics[width=\textwidth]{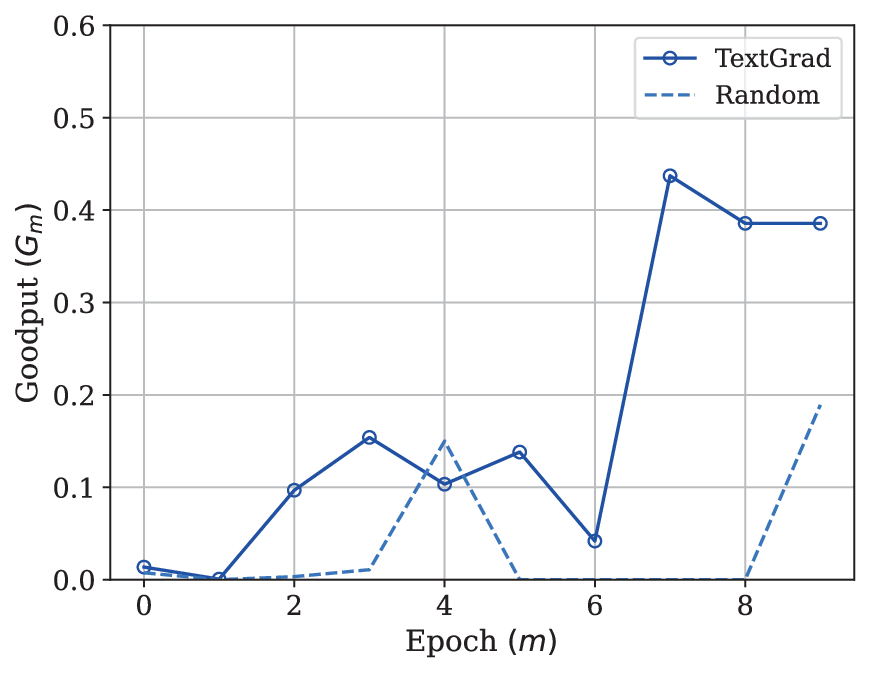}
    \caption{TPM w. TextGrad vs. random instructions.}
    \label{sim_fig:textgrad_vs_random}
    \end{subfigure}
    \hfill
    \begin{subfigure}{0.32\textwidth}
    \centering
    \includegraphics[width=\textwidth]{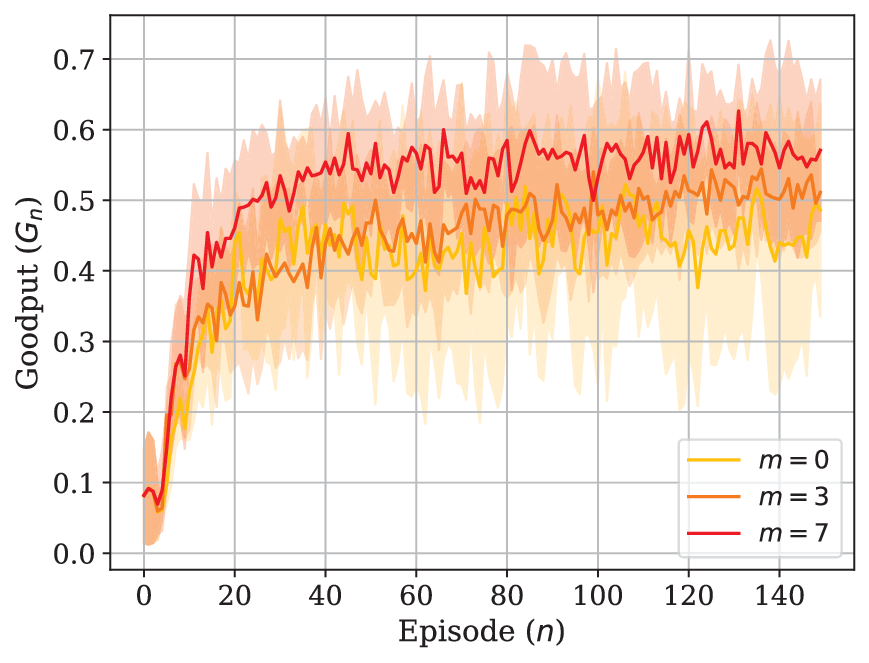}
    \caption{Goodput during re-training of T2NPM.}
\label{sim_fig:T2NPM_prompt}
    \end{subfigure}
    \hfill
    \begin{subfigure}{0.33\textwidth}
    \centering
    \includegraphics[width=\textwidth]{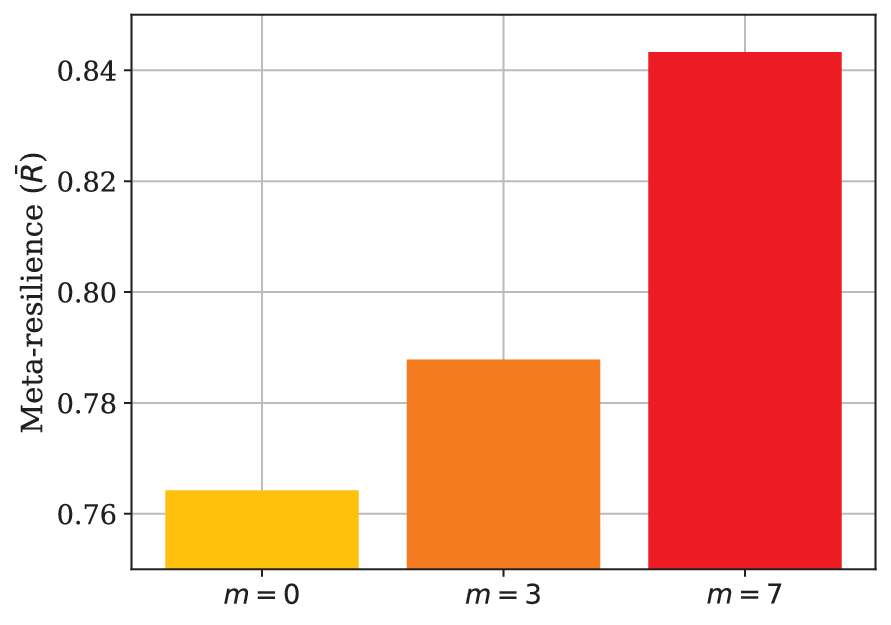}
    \caption{\centering Meta-resilience of T2NPM.}
\label{sim_fig:Meta-R T2NPM}
    \end{subfigure}
\caption{Impact of TextGrad on TPM and T2NPM with respect to the number $m$ of TextGrad epochs increases.}
\label{fig:T2NPM instructions}
\end{figure*}


\subsection{Impact of Environmental Shifts}
Table~\ref{tab: L increment validation} presents the goodput of NPM under various types of environmental shifts, where the NPM parameter set $\boldsymbol{\Theta}_0$ is pre-trained under the conditions $p_\ell^a = 0.3$, $b_\ell^\max = 3$, and $L = 2$. We then evaluate the following individual environmental shifts, with increases and decreases marked in red and blue, respectively: $p_{\ell}^a \in \{\textcolor{red}{0.5},\, \textcolor{blue}{0.1}\}$, $b_{\ell}^{\max} \in \{\textcolor{red}{5},\, \textcolor{blue}{2}\}$, $L \in \{\textcolor{red}{3},\, \textcolor{blue}{1}\}$, and $p_{\ell}^e \in \{\textcolor{red}{0.1},\, \textcolor{blue}{0.001}\}$. Two cases are considered,  \textcircled{\raisebox{-0.85pt}{1}}: goodput is evaluated using  $\boldsymbol{\Theta}_0$, without any adaptation to a new environment, \textcircled{\raisebox{-0.85pt}{2}}: $\boldsymbol{\Theta}_0$ is re-trained to adapt to a new environment. When \textcolor{red}{$L \uparrow$}, the fixed $\boldsymbol{\Theta}_0$ cannot accommodate the increased input/output dimension, requiring a randomly initialized NN to support the new UE. Among all evaluated environmental shifts, this scenario yields the largest performance gap between the original and re-trained NPMs. Therefore, we hereafter focus only on \textcolor{red}{$L\uparrow$}.

\subsection{Meta-resilience Analysis on TPM, T2NPM, and T3NPM}
Fig.~\ref{fig:numUE vs methods} illustrates the derivation from goodput to meta-resilience, highlighting the superior performance of T3NPM in terms of meta-resilience across all evaluated methods. The goodput $G_n$ over episodes is shown in Fig.~\ref{sim_fig:R triangle methods}. At the environmental shift episode $n_0$, both NPM (dotted red) and T2NPM (solid red) exhibit a significant drop in goodput since they cannot adapt to the sudden change in input-output dimensions caused by fixed NN architectures. In contrast, TPM (solid blue) maintains stable goodput, as it relies on natural language-based inference and remains independent of NN structural constraints. While NPM and T2NPM eventually recover from the performance drop, T2NPM trains faster and shows better goodput than NPM, benefiting from the proposed KD regularization. To guarantee that the T2NPM goodput set satisfies $|\mathcal{V}_n^{\text{T2NPM}}| = 12$, goodput is measured every 12 TTIs during the measurement phase, as in the testing phase. Since each goodput value requires 12 TTIs and we aim to collect 12 such samples, the total required measurement duration becomes $\hat{T}_M = 144$ TTIs. Given that each episode allocates $T_M = 24$ TTIs for measurement, only two goodput values can be obtained per episode. Therefore, $k = \lceil \hat{T}_M / T_M \rceil - 1 = 5$ previous episodes are utilized to construct $\mathcal{V}_n^{\text{T2NPM}}$. With this configuration, T3NPM avoids the severe performance degradation observed in NPM and T2NPM by initially operating as TPM. As training advances and T2NPM begins to outperform TPM, the protocol switches dynamically, enabling further learning and sustained long-term performance enhancement.

Since Fig.~\ref{sim_fig:R triangle methods} includes the effects of initial performance degradation, gradual recovery, and final convergence, it is difficult to directly evaluate the resilience of each method. Therefore, $R\left(\{G_n\}_{n\in \mathcal{N}}, \hat{G}\right)$ across various target goodput levels $\hat{G} \in [0.01, 1]$ is illustrated in Fig.~\ref{sim_fig:R methods}. When $\hat{G}$ is close to zero, the environmental shift does not cause any performance degradation in terms of resilience, resulting in all methods achieving $R\left(\{G_n\}_{n\in \mathcal{N}}, \hat{G}\right) = 1$. At low values of $\hat{G}$, both NPM and T2NPM exhibit lower resilience than TPM and S-ALOHA, primarily due to their significant performance degradation immediately following the environmental shift. However, as $\hat{G}$ increases beyond the goodput of TPM and S-ALOHA, their resilience sharply declines. T2NPM consistently outperforms NPM in resilience across all $\hat{G}$ values. T3NPM achieves the highest resilience over the entire range of $\hat{G}$, as it avoids the initial degradation through early-stage TPM deployment and then transitions to T2NPM for long-term performance gains. Because comparisons based on specific $\hat{G}$ values can fluctuate, often resulting in crossing points among the resilience curves, Fig.~\ref{sim_fig:Meta-R methods} summarizes the overall results using $\bar{R}\left(\{G_n\}_{n\in\mathcal{N}}\right)$, defined in \eqref{eq:meta resilience}.
For T3NPM meta-resilience calculation, $\bar{R}\left(\{G_n\}_{n \in \mathcal{N}} \mid 
T_M, k\right)$ in \eqref{p1:T3NPM opt} is used. Among the methods, T3NPM's $\bar{R}\left(\{G_n\}_{n\in\mathcal{N}}\mid 24, 5\right)$ achieves the highest meta-resilience, outperforming T2NPM, TPM, NPM, and S-ALOHA by $4.37\%$, $10.05\%$, $20.56\%$, and $23.53\%$, respectively. The rationale of deciding $T_M = 24$ and $k = 5$ is explained in Sec.~\ref{subsection:simulation MixSwitch}. Since meta-resilience accounts for performance across a broad spectrum of target levels, we conclude that T3NPM provides the most resilient performance among all proposed methods.

\subsection{Impact of TextGrad}
The impact of TextGrad on both TPM and T2NPM, specifically on their goodput and meta-resilience, is illustrated in Fig.~\ref{fig:T2NPM instructions}, highlighting the effect of instruction quality. Fig.~\ref{sim_fig:textgrad_vs_random} compares TPM performance under different instruction acquisition methods across training epochs $m \in [0, 9]$. In the `random', instructions are generated without any feedback. In contrast, the `TextGrad' approach iteratively refines instructions using LLM-generated feedback based on the objective function. At each epoch $m$, $G^{\phi_m}$ is evaluated using TPM with instruction $\phi = \phi_m$. The results demonstrate that instructions generated via TextGrad significantly outperform those acquired randomly. This improvement stems from the LLM’s capacity to comprehend the natural language expression of the task objective, generate meaningful feedback, and update the instruction accordingly. Due to the stochastic nature of LLM-generated feedback and updates, increasing the number of epochs does not always consistently improve instruction performance. Therefore, the highest TPM goodput is achieved among all evaluated instructions using $\phi = \phi_7$.

Since TPM is also utilized as the teacher model in T2NPM, the influence of instruction quality on T2NPM performance is illustrated in Fig.~\ref{sim_fig:T2NPM_prompt}. Here, instructions $\phi_0$, $\phi_3$, and $\phi_7$, obtained from Fig.~\ref{sim_fig:textgrad_vs_random}, are applied to the teacher model while training the student model. For clearer comparison, Fig.~\ref{sim_fig:Meta-R T2NPM} presents the corresponding meta-resilience results. As expected, instruction quality directly affects T2NPM performance. The meta-resilience $\bar{R}\left(\{G_n^{\phi_7}\}_{n \in \mathcal{N}}\right)$ exceeds $\bar{R}\left(\{G_n^{\phi_0}\}_{n \in \mathcal{N}}\right)$ and $\bar{R}\left(\{G_n^{\phi_3}\}_{n \in \mathcal{N}}\right)$ by $10.35\%$ and $7.04\%$, respectively. Although KD does not require an optimal teacher model, these results emphasize the critical role of instruction quality in enhancing the student model's goodput. Based on these findings, instruction $\phi = \phi_7$ is adopted for all simulations, including those in Fig.~\ref{fig:numUE vs methods}.

\begin{table}[t]
\captionsetup{justification=centering, labelsep=newline, font={smaller,sc}}
\caption{Computational cost at testing phase.}
\begin{center}
\begin{tabular}{l|c|c|c}
\toprule
  & \multicolumn{3}{c}{\textbf{Computational cost}} \\ 
\textbf{Scheme} & Avg. FLOPS & Parameters & Memory \\
\hline
TPM & 7.13T & 10.7B & 41.1GiB \\
T3NPM, $T_M = {120}$ & 6.05T & 10.7B & 41.3GiB \\
T3NPM, $T_M = {96}$ & 1.43T & 10.7B & 41.3GiB \\
T3NPM, $T_M = {72}$ & 0.90T & 10.7B & 41.3GiB \\
T3NPM, $T_M = {48}$ & 0.56T & 10.7B & 41.3GiB \\
T3NPM, $T_M = {24}$ & \textbf{0.36T} & 10.7B & 41.3GiB \\
T2NPM & 0.11M & 50.83K & 190MiB \\
NPM & 0.11M & 50.83K & 190MiB \\
\bottomrule
\end{tabular}
\label{tab:Model size comparison}
\end{center}
\end{table}

\subsection{Impact of KD}
KD enables T2NPM to achieve both faster convergence and higher convergence goodput than NPM, while maintaining significantly lower computational complexity than TPM, as illustrated in Fig.~\ref{sim_fig:R triangle methods} and  the first three rows in Table~\ref{tab:Model size comparison}. As shown in Fig.~\ref{sim_fig:R triangle methods}, T2NPM demonstrates a steeper increase in goodput during early episodes compared to NPM. This improvement results from the KD regularization term, which guides the student model to imitate the teacher model’s actions, thereby simplifying the training process, particularly in the early stages where learning from scratch is challenging. Over time, T2NPM surpasses both NPM and TPM in terms of convergence goodput. This performance gain stems from the combined effect of the KD loss, which promotes policy alignment with TPM, and the TD loss, which allows the model to interact with the environment and refine its policy. In terms of computational cost, Table~\ref{tab:Model size comparison} confirms that T2NPM and NPM incur equivalent costs, which are significantly lower than that of TPM. Average FLOPS is obtained by dividing the total FLOPs across all episodes by $|\mathcal{N}|$. Due to their lightweight architectures with fewer parameters and lower memory requirements, both NPM and T2NPM exhibit minimal FLOPs per episode. In contrast, TPM incurs the highest computational cost, as it requires inference using an LLM for all episodes, resulting in substantially greater FLOPs.

\begin{figure}[t]
\centering
    \begin{subfigure}{0.9\columnwidth}
        \centering \includegraphics[width=0.85\columnwidth]{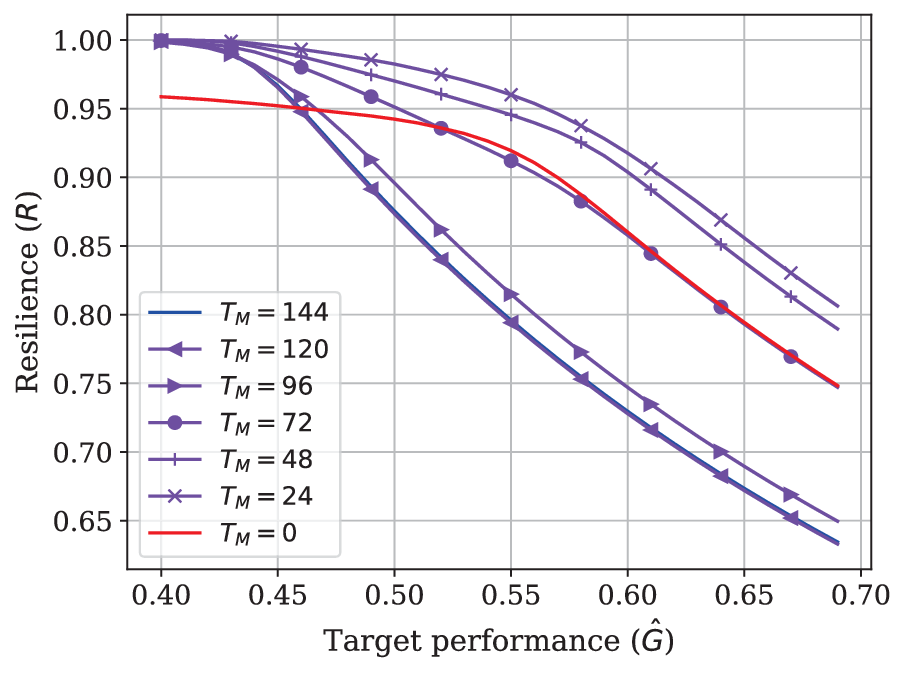}
        \caption{Resilience.}
        \label{sim_fig:R_T3NPM}
    \end{subfigure}
    \vspace{0.1em} 
    \begin{subfigure}{0.9\columnwidth}
        \centering
        \includegraphics[width=0.85\columnwidth]{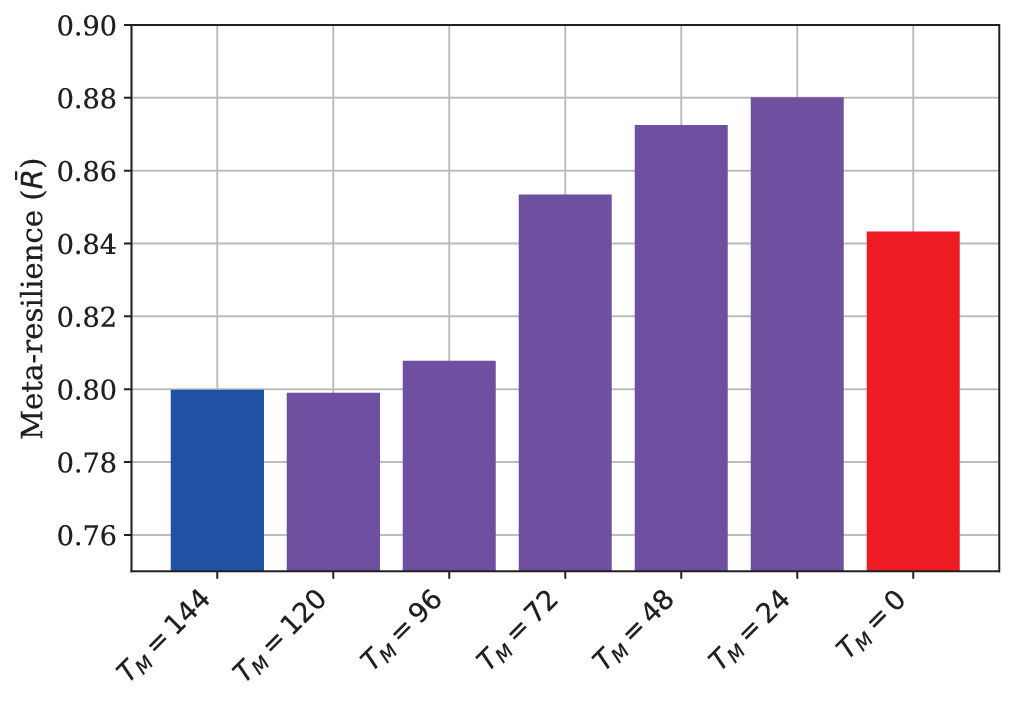}
        \caption{Meta-resilience.}
        \label{sim_fig:MetaR_T3NPM}
    \end{subfigure}
    \caption{Resilience and meta-resilience for different $T_M$'s.}
     \label{fig:T3NPM_configurations}
\end{figure}

\subsection{Impact of MixSwitch}
\label{subsection:simulation MixSwitch}
The meta-resilience of T3NPM varies with the choice of measurement phase duration $T_M$, as it simultaneously determines the training time ($T - T_M$) and the number of past episodes $k$ used in the MixSwitch. Therefore, solving the optimization problem in \eqref{p1:T3NPM opt} is required to identify the optimal $\hat{T}_M$. Fig.~\ref{fig:T3NPM_configurations} compares T3NPM’s resilience and meta-resilience across multiple values of $T_M$. To ensure a consistent sample size $|\mathcal{V}_n^{\text{T2NPM}}| = 12$, the feasible set of $T_M$ is constrained to values divisible by $12$ TTIs as $T_M \in \{0, 24, 48, 72, 96, 120, 144\}$, and T3NPM with each configuration is denoted as $\text{T3NPM}_{T_M}$. Resilience for each configuration is illustrated in Fig.~\ref{sim_fig:R_T3NPM}, while Fig.~\ref{sim_fig:MetaR_T3NPM} presents the corresponding meta-resilience values. Notably, $T_M = 144$ and $T_M = 0$ correspond to TPM and T2NPM, respectively. As $T_M$ decreases from $144$ to $24$, more training time is allocated per episode, improving long-term performance. However, when $T_M = 0$, no measurement phase is performed, requiring the system to operate as T2NPM from the beginning, which limits adaptability and reduces meta-resilience. According to Fig.~\ref{sim_fig:MetaR_T3NPM}, the configuration $T_M = 24$ achieves the highest meta-resilience, with gains of $10.16\%$, $8.96\%$, $3.13\%$, and $0.87\%$ over $T_M = 120$, $T_M = 96$, $T_M = 72$, and $T_M = 48$, respectively. Consequently, $T_M = 24$ is selected as the optimal solution to \eqref{p1:T3NPM opt}, and this configuration is adopted for all simulations, including Fig.~\ref{fig:numUE vs methods}. Additionally, a smaller $T_M$ results in rapid goodput improvement, thus an earlier switching point from TPM to T2NPM. It further reduces average FLOPS during the testing phase, as shown in Table~\ref{tab:Model size comparison}.

\begin{table}[t]
\captionsetup{justification=centering, labelsep=newline, font={smaller,sc}}
\caption{Impact of ICL under environmental shifts.}
\begin{center}
\begin{tabular}{c|c|c|c|c|c|c}
\toprule
\textbf{Environmental} & \multicolumn{6}{c}{\textbf{Number of few-shot examples}} \\ 
\textbf{Shift} & 0 & 2 & 4 & 6 & 8 & AGR \\
\hline
$L \uparrow$ & 0.44 & 0.61 & 0.57 & 0.61 & 0.60 & 9.36\%\\
$p_\ell^a \uparrow$ & 0.41 & 0.58 & 0.62 & 0.68 & 0.73& 16.35\%\\
$b_\ell^{\max} \uparrow$ & 0.21 & 0.44 & 0.41 & 0.45 & 0.45& 28.12\%\\
$p_\ell^a \uparrow, b_\ell^\max  \uparrow$ & 0.28 & 0.57 & 0.61 & 0.67 & 0.71 & 31.60\%\\
\bottomrule
\end{tabular}
\label{tab:ICL}
\end{center}
\end{table}

\subsection{Ablation Studies: Impact of ICL and SNR} \label{Sec:Ablation}

As discussed in Sec.~\ref{subsection:TextGrad-TPM}, ICL and prompt engineering via TextGrad are two promising LLM adaptation methods for TPM construction. However, ICL is effective only under mild environmental shifts, as it relies on exemplary queries that become irrelevant under sever shifts. To illustrate, Tab.~\ref{tab:ICL} reports the average goodput of TPM with ICL using $m=2$ examples across four types of environmental shifts: 1) $L \uparrow$, 2) $p_{\ell}^a \uparrow$, 3) $b_\ell^{\max} \uparrow$, and 4) joint shift in $p_\ell^a \uparrow$ and $b_\ell^\max \uparrow$. In this setting, only pre-shift data are available for constructing ICL examples. The results show that ICL remains effective under mild environmental shifts involving 2)-4), where pre-shift examples are still relevant. In contrast, under the severe shift 1), ICL yields little performance gain, in terms of the average growth rate (AGR) of goodput. Based on these findings, we recommend using TextGrad alone when adapting to severe environmental shifts like $L \uparrow$, while combining ICL with TextGrad is effective for mild shifts such as $p_\ell^a$ and/or $b_\ell^\max$.

\begin{figure}[t]\centering
\includegraphics[width=0.76\columnwidth]{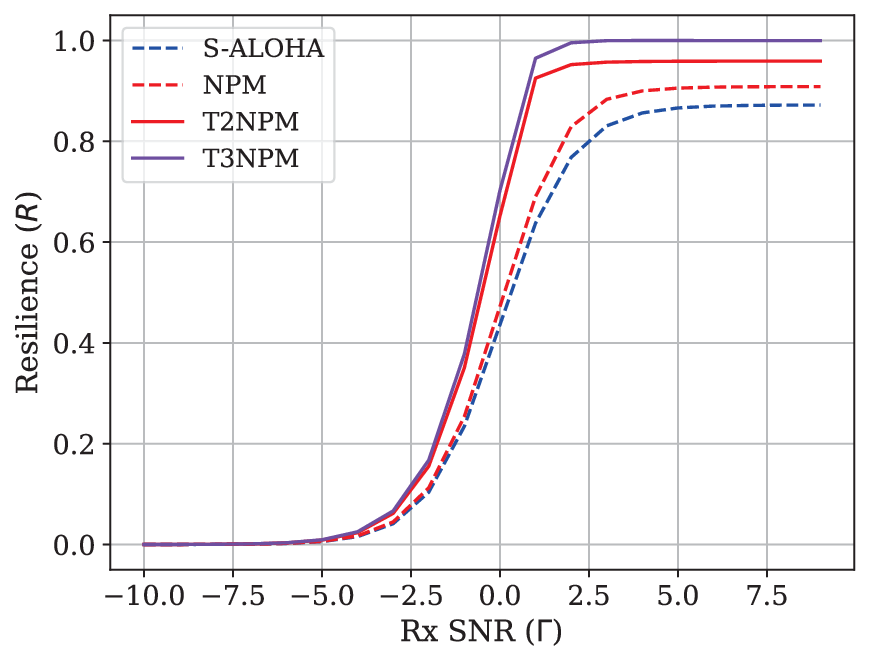}
\caption{\centering Impact of SNR on resilience.}
\label{fig:bler_goodput}
\end{figure}


Next, we examine the case of reduced BLER $p_\ell^e \downarrow$, under the concurrent $L\uparrow$ shift. The resulting loss in resilience can be compensated by re-training as shown in Tab.~\ref{tab: L increment validation}, or alternatively by increasing transmit power or equivalently signal-to-noise ratio (SNR). To illustrate the latter, for a fixed target goodput level $\hat{G}=0.4$, Fig.~\ref{fig:bler_goodput} depicts resilience as SNR increases, where BLER is modeled as a sigmoid function of SNR~\cite{Carreras:2018}. T3NPM consistently achieves the highest resilience and gains the most from increasing SNR. Consequently, to reach a resilience of $R=0.8$, T3NPM requires $2\,\mathrm{dB}$ and $0.7\,\mathrm{dB}$ lower SNR compared to S-ALOHA and NPM, respectively. Nevertheless, resilience saturates at high SNR in all methods, leaving non-negliglble resilience gaps that highlight the need for improved protocol designs. Notably, T3NPM only achieves the maximum $R=1$, owing to its ability to meet $\hat{g}$ using TPM immediately after the shift, underscoring the important role of LLMs in resilient protocol designs.


\section{Conclusion}
\label{section:conclusion}
In this paper, we addressed the challenges posed by environmental shifts in NPM, which require re-training and thus compromise resilience. To overcome this, we proposed LLM-based semantic MAC protocols: TPM, T2NPM, and T3NPM. Leveraging LLM-based inference, TPM immediately enables coarse-yet-acceptable MAC operations under environmental shifts, facilitated by LLM-based instruction fine-tuning through TextGrad, without the need for re-training. By utilizing TPM as a coarse teacher, T2NPM not only transfers the knowledge of its LLM into a small NN but also accelerates the NN's re-training via KD, thereby reducing inference costs and improving goodput. Finally, T3NPM combines TPM for immediate responses and T2NPM for reduced computation and enhanced goodput, using MixSwitch to optimize the TPM-to-T2NPM transition point. To define the objective for this transition optimization and enable fair comparison across different methods, we introduced a novel metric, meta-resilience, which quantifies resilience when target performance is undefined, a crucial aspect for capturing the unknown impacts of environmental shifts. Simulations demonstrated that T3NPM outperforms NPM, TPM, T2NPM, and S-ALOHA in meta-resilience, under environmental shifts caused by variations in the number of UEs and SNR. Building on this, a promising direction for future study is to evaluate the effectiveness of LLM-based semantic MAC protocols in more realistic MAC scenarios. Furthermore, exploring the additional role of LLM's reasoning capabilities during test-time compute could be an interesting future topic.


\bibliographystyle{IEEEtran}
\bibliography{references}

\end{document}